\documentclass[twocolumn,aps,prb,nofootinbib]{revtex4-1}
\usepackage{graphicx}
\usepackage{times}
\usepackage{bm}
\usepackage[linktocpage=true,colorlinks=true,urlcolor=blue,linkcolor=red,citecolor = blue]{hyperref}
\usepackage{pgfplots}
\usepackage{amsfonts}
\usepackage{amsmath}
\usepackage{xcolor}
\usepackage{braket}
\usepackage{relsize}
\usetikzlibrary{arrows}
\usepackage{enumerate}
\usetikzlibrary[patterns] 
\usetikzlibrary[snakes] 
\usetikzlibrary[shapes] 
\usepackage{pgfplots}
\usepackage{newfloat}
\usepackage{float}

\DeclareFloatingEnvironment[name={Supplementary Figure}]{suppfigure}

\graphicspath{ {Figures/} }

\begin{document}

\title{Enhanced topological protection in planar quasi-one-dimensional channels with periodically-modulated width}

\author{Benjamin D. Woods}
\affiliation{Department of Physics and Astronomy, West Virginia University, Morgantown, WV 26506, USA}
\author{Tudor D. Stanescu}
\affiliation{Department of Physics and Astronomy, West Virginia University, Morgantown, WV 26506, USA}

\begin{abstract}
We study one dimensional (1D) and quasi-1D periodic structures as possible  platforms for the emergence of Majorana bound states with enhanced robustness against disorder and system inhomogeneity. First, using a simple 1D model, we analytically derive the effective parameters characterizing the minibands generated by the periodic potential. We show that,  for strong enough periodic potentials, the higher energy minibands hosting Majorana bound states have significant  advantages compared to their counterparts in uniform systems, including increased topological gaps, enhanced robustness against disorder, and enlarged parameter space regions consistent with the presence of topological superconductivity. We identify the problem of engineering a strong enough periodic potential as a key roadblock to realizing efficient periodic 1D structures. To address this challenge, we propose an efficient  implementation of the periodic potential based on quasi-1D channels realized in 2D semiconductor heterostructures proximity coupled to superconductor strips of periodically modulated width. Our numerical study of the modulated channel device shows excellent agreement with the simple 1D model, reveals a topological phase diagram that is quite insensitive to the details of the confining potential associated with screening by the superconductor, and demonstrates that engineering patterned 2D structures represents a powerful and versatile approach to realizing robust Majorana bound states. 
\end{abstract}

\maketitle

\section{Introduction}
Majorana bound states (MBSs) -- also called Majorana zero modes (MZMs) -- are zero-energy modes localized near topological defects in topological superconductors, e.g., at vortex cores in a 2D system or near the edges of a 1D system \cite{Bravyi2002}. These states have attracted much attention in the last decade due to their topological protection and non-Abelian statistics, which make them promising candidates as a platform for implementing topological quantum computing \cite{Nayak2008,Alicea2011,Lahtinen2017}. Following theoretical proposals  for realizing topological superconductivity using spin-orbit coupled semiconductors and s-wave superconductors \cite{Lutchyn2010,Oreg2010}, significant experimental effort has been dedicated to the search for signatures of zero-energy MBSs in proximity-coupled semiconductor-superconductor (SM-SC) heterostructures \cite{Mourik2012,Deng2012,Das2012,Rokhinson2012,Churchill2013,Finck2013,Albrecht2016,Deng2016,Zhang2017,Chen2017,Nichele2017,Zhang2018,Lutchyn2018}. While many observations have been consistent with the predicted MBS signatures, there is a persistent concern that at least some (and possibly all) of the features observed so far are, in fact, due to topologically trivial Andreev bound states (ABSs) that mimic the Majorana phenomenology \cite{Kells2012,Moore2016,Moore2018,Stanescu2018b,Pan2019}. These low-energy Andreev bound states can be induced through several mechanisms, including soft confinement \cite{Kells2012,Stanescu2018b,Vuik2019}, disorder \cite{Bagrets2012,Liu2012,Adagideli2014,Pan2019}, inhomogeneous superconductivity \cite{Roy2013,Cayao2015,Fleckenstein2018}, nonuniform spin-orbit coupling \cite{Cao2019}, and inter-subband coupling \cite{Woods2019b}. In essence, the root cause for the emergence of these (topologically trivial) low-energy ABSs is the presence of some type of non-uniformity within the system. Accessing the topological phase would require applying a strong-enough magnetic field corresponding to a Zeeman splitting larger than the characteristic energy scale associated with the non-uniformity, which may not be possible without completely destroying the superconducting gap. The emergence of  low-energy ABSs mimicking the Majorana phenomenology at low values of the magnetic field, which is well established theoretically \cite{Kells2012,Liu2012,Roy2013,Cayao2015,Moore2016,Dominguez2016,Moore2018,Avila2018,Stanescu2018b,Pan2019,Woods2019b}, can be viewed as a precursor to Majorana physics, in the sense that topological superconductivity will necessarily emerge if the system inhomogeneity is reduced, or if the magnetic field is enhanced without destroying   superconductivity.  However, realizing MBSs that are practically relevant for quantum computing requires devices that reliably give rise to well separated Majorana modes and are free of topologically-trivial low-energy ABSs. While reducing the inhomogeneities within the system below an acceptable level may prove difficult, one can adopt a completely different strategy and amplify the non-homogeneity in a controlled manner by applying a periodic potential, i.e. creating a superlattice \cite{Adagideli2014,Levine2017,Lu2016,Escribano2019}. 

In this work, we analytically and numerically study the emergence of MBSs within periodic structures and show that the periodic design alleviates many of the issues alluded to above.  We also propose an efficient design for realizing strong periodic potentials using quasi-1D  modulated-width channels in 2D semiconductor-superconductor heterostructures. We show that  periodic structures provide three main advantages when compared to uniform systems. (1) The total area of the parameter space associated with the topological superconducting phase increases. As pointed out in previous studies \cite{Adagideli2014,Levine2017,Lu2016,Escribano2019}, this is related to the formation of minibands in the presence of a periodic potential. (2) The topological gap characterizing the topological superconducting phase is typically larger than the corresponding value in a uniform system and, generally, increases with increasing chemical potential. In essence, this occurs as a result  of an increased effective spin-orbit coupling within the higher energy minibands. (3) The topological state shows increased robustness against disorder. This is an effect of the increased topological gap combined with the highly oscillatory nature of the states associated with  high-energy minibands.

In Sec. \ref{1D}, we explore the effects of a periodic potential within a purely 1D system based on the minimal model of a Majorana nanowire \cite{Lutchyn2010,Oreg2010}, with the addition of a periodic potential. Starting from a suitable low-energy basis, we analytically derive effective parameters characterizing each set of periodicity-induced minibands. These parameters include renormalized effective masses, spin-orbit coupling coefficients, Zeeman splittings, and Majorana localization lengths. The derived  analytic expressions provide valuable insight into the underlying physics and suggest possible avenues for optimizing the topological properties of the system. 
Numerical calculations based on an equivalent 1D tight-binding model show excellent agreement with the analytic results and highlight the importance of creating a sufficiently strong periodic potential, without which the periodic structure loses its advantages over the uniform system. The  formation of (topologically trivial) partially-separated Andreev bound states (ps-ABSs) \cite{Moore2018,Stanescu2018b,Vuik2019} in the presence of a soft confining potential is explored in Sec. \ref{Soft}. We find that the confinement must be softer (i.e. have a smaller slope) within the periodic system, as compared to a uniform one, to ensure the ps-ABS collapse to zero-energy. Thus, the presence of a periodic potential reduces the parameter region associated with the presence of ps-ABSs, which may provide a significant advantage in the search for topologically protected MBSs.  In addition, we investigate the effects of potential disorder (see Sec. \ref{DIS_1D}) and find that, typically,  the topological phase becomes more robust in the presence of an additional periodic potential. Moreover, in a superlattice the nonlocal (edge-to-edge) correlations indicative of well-separated MBSs localized at the ends of the system are found to become less sensitive to the presence of disorder. 

While a periodic potential can, in principle, offer the advantages mentioned above, engineering a strong-enough potential represents a nontrivial task. Naively, one could try to generate such a potential using  periodic arrays of gates applied to ``standard'' semiconductor wire-superconductor devices similar to those used in recent Majorana experiments \cite{Mourik2012,Krogstrup2015,Gazibegovic2017,Kang2017,Sestoft2018,Krizek2019}. 
We find that, unfortunately, the effective periodic potential generated within such a nanowire setup is too weak for the superlattice scheme to provide any notable advantage over the uniform system.
A possible alternative is discussed in Sec. \ref{Waveguide}. The proposed device (see Fig. \ref{FIG_2_1}), which we will refer to as a \textit{modulated channel device} or a \textit{Majorana waveguide}, consists of a two-dimensional electron gas (2DEG) hosted by a semiconductor heterostructure and proximity-coupled to a lithographycally defined superconductor that generates a quasi-1D channel with periodically modulated width. The feasibility of this type of structure is supported by the recent progress in fabricating two-dimensional epitaxial superconductor-semiconductor heterostructures that support low-energy features similar to those observed in proximitized wires \cite{Shabani2016,Suominen2017,Nichele2017,OFarrell2018,Fornieri2019,Lee2019,Mayer2019}. While a periodic potential is not directly applied, the periodic structure of the device results in the formation of minibands similar to those induced by an actual periodic potential, due to scattering at the interfaces between regions of differing width \cite{Weisshaar1991}. 
For not-too-large values of the chemical potential, the topological properties of the system, including the stability of the MBSs, exhibit all the advantageous features identified in the 1D ``ideal'' periodic model. In addition, we show that the topological  phase diagram is not dramatically affected by the details associated with the screening by the superconductor of the confinement potential that defines the modulated channel. Specifically, we consider a confining potential that varies near the edges of the region covered by the superconductor over a finite length scale $\chi$ and show that the phase diagram depends weakly on $\chi$. Furthermore, we argue that $\chi$ can be quite large, which implies  that the chemical potential of the electron gas underneath the superconductor is tunable (to a certain degree), providing an important knob for accessing the topological superconducting phase. 
Taking into account all these findings, as well as the natural ability of the 2DEG system to enable the construction of complex structures, we conclude that the {Majorana waveguide} and, more generally,  patterned 2D structures represent a promising versatile platform for realizing robust Majorana bound states.

\section{One-dimensional model of periodic Majorana nanowires} \label{1D} 

In this section we study analytically and numerically a simple, one-dimensional model of a Majorana wire in the presence of a periodic effective potential. We show that topological superconductivity and Majorana zero modes emerge at low values of the applied Zeeman field (on the order of the induced pairing potential) whenever the chemical potential is tuned near the bottom/top of a potential-induced pair of minibands. We determine explicit analytical expressions for the renormalized miniband parameters (e.g., effective mass and spin-orbit coupling) and the Majorana localization length in periodic nanowires. The validity of these analytic expressions is verified numerically. We also investigate the effect of the periodic potential on the topological phase diagram, the emergence of topologically trivial Andreev bound states in systems with soft confinement, and the robustness of Majorana bound states against disorder. 

\subsection{Renormalization of miniband parameters}

\begin{figure}[t]
\begin{center}
\includegraphics[width=0.48\textwidth]{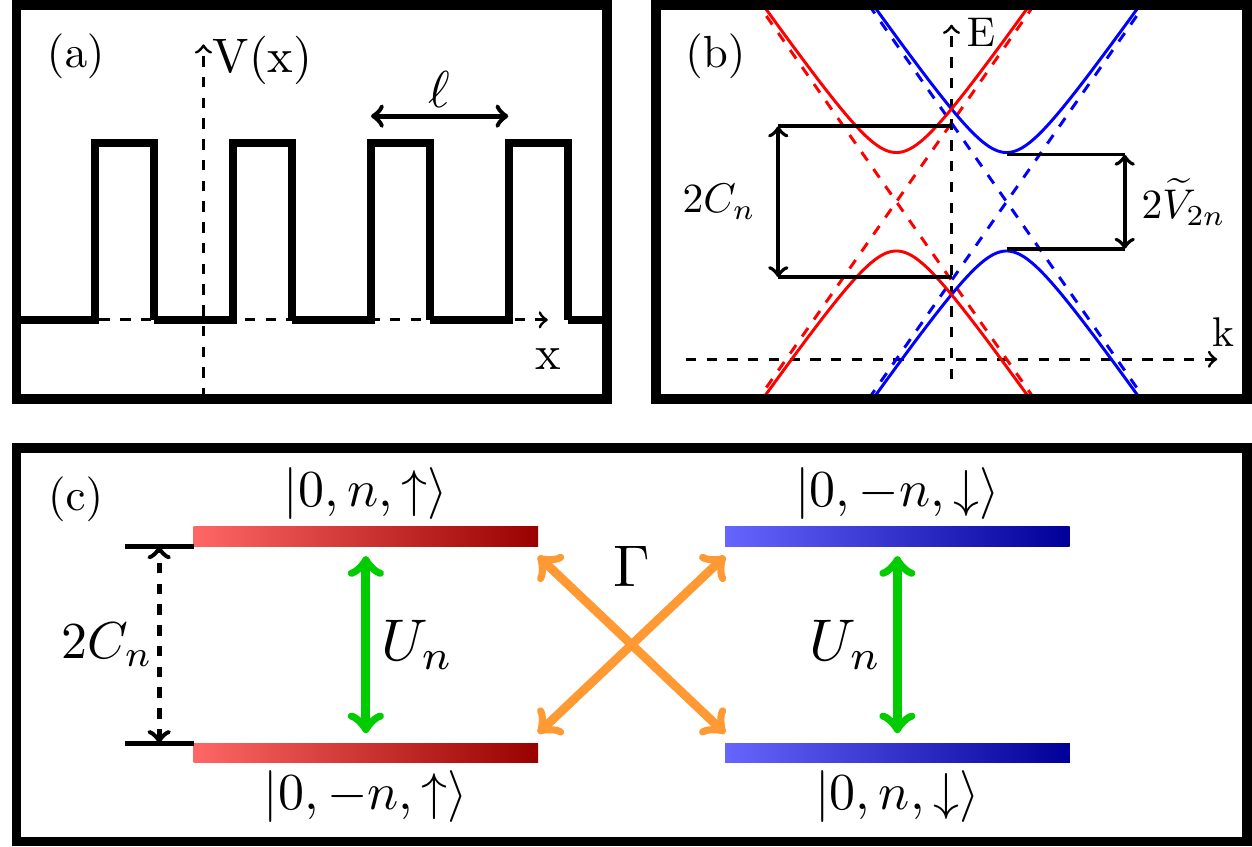}
\end{center}
\vspace{-5mm}
\caption{(a) Rectangular 1D potential profile with period $\ell$. (b) Spectrum of the effective Hamiltonian (\ref{Hn}) near the zone center for zero magnetic field ($\Gamma = 0$) in the absence (dashed lines) and presence (solid lines) of a periodic potential. The red (blue) lines  correspond to spin $\sigma = \uparrow \left(\downarrow\right)$. (c) Energy level diagram showing the couplings of the zone $n$ basis states with $k = 0$ induced by the periodic potential and the applied magnetic field  The color code is the same as in (c).}
\label{FIG1}
\vspace{-3mm}
\end{figure}

We begin by considering  a minimal model of the Majorana nanowire \cite{Lutchyn2010,Oreg2010} in the presence of a periodic potential. 
More specifically, we have a one-dimensional system (i.e., a Majorana wire) with Rashba spin-orbit coupling, (induced) superconductivity,  magnetic field applied parralel to the wire, and a periodic (effective) potential. The system  is modeled by the Bogliubov-de Gennes  (BdG) Hamiltonian
\begin{equation}
H_{BdG}(k) = 
\begin{pmatrix}
    H_o\left(k\right) & -i\Delta\sigma_y \\
    i \Delta^* \sigma_y & -H_o^*\left(-k\right)
\end{pmatrix}. \label{HamBdG}
\end{equation}
The diagonal (normal wire) component is
\begin{equation}
    H_o(k) = \frac{\hbar^2 k^2}{2 m^*} - \mu + V\left(x\right)+ \alpha k \sigma_z + \Gamma \sigma_x, \label{Ho}
\end{equation}
where $m^*$ is the effective mass, $\mu$ is the chemical potential, $\alpha$ is the Rashba spin-orbit coefficient, $\Gamma$ is the (half) Zeeman splitting, and $\Delta$ is the (induced) superconducting pairing. Note that $\sigma_i$ with $i = x,y,z$ are the Pauli matrices acting within the spin space. The potential $V\left(x\right)$ is periodic with the period $\ell$, so that $V\left(x + \ell\right) = V\left(x\right)$.  A specific example of a periodic potential used in the calculations  is shown in Fig. \ref{FIG1}(a). We first study the quantum problem described by the BdG Hamiltonian with periodic boundary conditions, so that $k$ becomes a good quantum number. In the absence of the magnetic field and periodic potential, the normal Hamiltonian eigenstates are simply given by the polarized plane waves
\begin{equation}
\left<x\right|\left.k,n,\sigma\right> = \frac{\chi_\sigma}{\sqrt{L}}  \label{states}
e^{i\left(k+G_n\right)x},
\end{equation}
where $L$ is the total length of the system, $\chi_\sigma = \left(\delta_{\sigma,\uparrow},\delta_{\sigma,\downarrow}\right)^T$, $G_n = 2\pi n / \ell$ is a reciprocal (super)lattice vector, and k is restricted to the first Brillioun zone, $-\pi/\ell < k \leq \pi/\ell$. The label $n$ is the zone number of the plane wave state corresponding to the (superlattice) Brillioun zone to which the state belongs.  The energies of the eigenstates (\ref{states}) are given by
\begin{equation}
    E_{k,n,\sigma} = \frac{\hbar^2}{2m^*} \left(k + G_n\right)^2 + \alpha \left(k +  G_n\right)\left(\sigma_z\right)_{\sigma\sigma} - \mu. 
\end{equation}
To understand the effects of the periodic potential and magnetic field, it is convenient to Fourier transform $V(x)$ and to calculate the matrix elements of the pertubations (i.e., periodic potential and magnetic field) in the  plane wave basis given by Eq. (\ref{states}). We have
\begin{eqnarray}
    &~&V\left(x\right) = \sum_{n=-\infty}^{\infty}\widetilde{V}_n e^{-iG_nx},\\ \label {VFourier}
    &~&\left<k,m,\sigma| V\left(x\right)| k,n,\sigma^\prime \right> = ~\widetilde{V}_{n-m}\delta_{\sigma,\sigma^\prime}, \\ 
    &~&\left<k,m,\sigma| \Gamma\sigma_x| k,n,\sigma^\prime \right> =\Gamma\delta_{m,n}\left(\sigma_x\right)_{\sigma \sigma^\prime}.
\end{eqnarray}
Provided the period $\ell$ is sufficiently small,  the energy difference between different minibands, i.e. the difference between $E_{k,m,\sigma}$ and $E_{k,n,\sigma^\prime}$ with $|m| \neq |n|$, is much larger than the characteristic energy scales associated with the periodic potential and the magnetic field. Consequently, the basic physics can be understood by treating $V(x)$ and $\Gamma$ as perturbations acting within the subspace of basis states having the same absolute value of the zone number, i.e. $\left\{ \left|k,n,\uparrow\right>,\left|k,-n,\uparrow\right>,\left|k,n,\downarrow\right>,\left|k,-n,\downarrow\right>\right\}$.
For $n \neq 0$, the effective normal state Hamiltonian acting within this subspace is
\begin{equation}
\begin{aligned}
    H_{n}\left(k\right) =& 
\begin{pmatrix}
    C_n + v_+ k & U_n^* & \Gamma & 0 \\
    U_n & -C_n - v_- k & 0 & \Gamma \\
    \Gamma & 0 & -C_n + v_- k & U_n^* \\
    0 & \Gamma & U_n & C_n - v_+ k
\end{pmatrix}  
\\ &-\mu_n + \frac{\hbar^2 k^2}{2m^*}, \label{Hn}
\end{aligned}
\end{equation}
where the effective parameters are 
\begin{align}
\mu_n &= \mu - \frac{2 n^2 \pi^2 \hbar^2}{m^* \ell^2} - \widetilde{V}_0,\\
C_n &= \alpha \frac{2 \pi n}{\ell},\\
v_{n\pm} &= n \frac{2 \pi \hbar^2}{m^* \ell} \pm \alpha, \\
U_n &= V_{2n}.
\end{align}
Notice that for $\Gamma = 0$ (i.e., no magnetic field), spin is a good quantum number [i.e., the red and blue bands in Fig. \ref{FIG1} (b) and (c) do not couple] and one can exactly diagonalize the Hamiltonian in Eq. (\ref{Hn}). The energies of the corresponding eigenstates are
\begin{equation}
\begin{aligned}
\varepsilon_{\pm,\sigma}^{\left(n\right)} \left(k\right)=& 
\pm\sqrt{ \left(C_n + \left(\sigma_z\right)_{\sigma \sigma} \bar{v}_n k\right)^2 + |U_n|^2 } \\
&+ \left(\sigma_z\right)_{\sigma \sigma} \alpha k - \mu_n + 
\frac{\hbar^2 k^2}{2m^*},
\end{aligned}
\end{equation}
with $\bar{v}_n = \frac{1}{2} \left(v_{n+} + v_{n-}\right)$. The spectrum of a typical miniband in the vicinity of $k=0$  is shown in Fig. \ref{FIG1}(b). When there is no periodic potential ($U_n = 0$, dashed lines), the spectrum consists of  two Dirac-like cones with intersections shifted away from $k = 0$ due to the spin-orbit coupling. This gives rise to an energy splitting $2~C_n$ at $k = 0$, i.e. a splitting proportional to the Rashba coefficient ($\alpha$) and the miniband index ($n$). Applying a periodic potential opens a gap of size $2\left|U_n\right|$  at the nodes of the Dirac cones. 
Expanding the eigenenergies $\varepsilon_{\pm,\sigma}^{\left(n\right)}$ in the wave vector near $k = 0$, we have
\begin{equation}
    \varepsilon_{\pm,\sigma}^{\left(n\right)} \approx 
    \frac{\hbar^2 k^2}{2 m^*_{n\pm}}
    + \left(\sigma_z\right)_{\sigma\sigma} \alpha_{n\pm}k
    - \widetilde{\mu}_{n\pm}, \label{NRG}
\end{equation}
where the effective spin-orbit coupling $\alpha_{n\pm}$, mass $m^*_{n\pm}$, and chemical potential $\widetilde{\mu}_{n\pm}$ are re-normalized by the periodic potential, 
\begin{eqnarray}
    \alpha_{n\pm} &=& \left[1 \pm \frac{4\pi^2 n^2}{\gamma_n} \left(\frac{\hbar^2}{|U_n|m^*\ell^2}\right) \right]\alpha, \label{ALP} \\
    \frac{1}{m^*_{n\pm}} &=& \left[ 1 \pm  \frac{4\pi^2 \hbar^2 n^2}{m^* \ell^2 |U_n| \gamma_n}\left(1 \mp \left|\frac{C_n}{U_n \gamma_n}\right|^2     \right) \right]\frac{1}{m^*}, \label{MEFF}  \\
    \widetilde{\mu}_{n\pm} &=&  \mu_n  \mp \sqrt{C_n^2 + |U_n|^2}, \label{MUnT}
\end{eqnarray}
with $\gamma_n =    \sqrt{1 + \left|{C_n}/{U_n}\right|^2}$.
Remarkably,  both the upper and lower pairs of minibands mimic the spectrum of a uniform Rashba nanowire, but having renormalized effective masses and spin-orbit parameters that depend on the characteristics (i.e., amplitude and period) of the periodic potential. This suggests the possibility of optimizing the effective parameters of the nanowire by engineering the periodic potential. For example, Eq. (\ref{ALP}) shows  that, for moderate values of $n$, the renormalized spin-orbit coefficient can be significantly larger than the corresponding bare parameter $\alpha$. Combining Eqs. (\ref{ALP}) and (\ref{MEFF}), we obtain the effective spin-orbit energy
\begin{equation} 
    \widetilde{E}_{SO,n} = \pm \frac{m^* \alpha^2}{2 \hbar^2} 
    + \frac{2\pi^2 n^2}{\gamma_n}\left(\frac{\alpha^2}{|U_n|\ell^2}\right), \label{ESO}
\end{equation}
where the first term is the bare spin-orbit energy of the original Hamiltonian (\ref{Ho}) with no periodic potential, i.e. with $V(x)=0$, and the second term is a potential-induced contribution. Note that this additional contribution increases with the miniband index ($n$) and can become dominant, as we explicitly show below.

Next, we apply a magnetic field, $\Gamma \neq 0$, which removes the spin-degeneracy at $k=0$. For convenience we incorporate the effects of the magnetic field by writing the Hamiltonian (\ref{Hn}) in the basis of eigenstates corresponding to $\Gamma = 0$. Explicitly, we have  
\begin{equation}
\widetilde{H}_{n}(k) = 
\begin{pmatrix}
\varepsilon_{+,\uparrow}^{\left(n\right)} & \widetilde{\Gamma} & \Omega & 0 \\
 \widetilde{\Gamma} & \varepsilon_{+,\downarrow}^{\left(n\right)} & 0 & -\Omega \\
\Omega  & 0 & \varepsilon_{-,\downarrow}^{\left(n\right)} &  \widetilde{\Gamma} \\
0 &  -\Omega  & \widetilde{\Gamma} & \varepsilon_{-,\uparrow}^{\left(n\right)}
\end{pmatrix}, \label{Hn2}   
\end{equation}
where $\widetilde{\Gamma}\left(k\right)$ and $\Omega\left(k\right)$ are intra- and inter-miniband pair coupling terms, respectively. 
Focusing on $k = 0$ and expanding $\widetilde{\Gamma}$ and $\Omega$ in a power series with respect to the parameter $\left(C_n / U_n\right)$, we have
\begin{equation}
    \widetilde{\Gamma}\left(k=0\right) = 
    \left(1 
    -\frac{1}{2}\left(\frac{C_n}{U_n}\right)^2  
    +\frac{3}{8}\left(\frac{C_n}{U_n}\right)^4
    +\mathcal{O}\left(\frac{C_n}{U_n}\right)^6
    \right) \Gamma, \label{RGam}
\end{equation}
\begin{equation}
    \Omega\left(k=0\right) = 
    \left( 
   \left(\frac{C_n}{U_n}\right)  
    -\frac{1}{2}\left(\frac{C_n}{U_n}\right)^3
    +\mathcal{O}\left(\frac{C_n}{U_n}\right)^5
    \right) \Gamma,  \label{Omega}
\end{equation}
where, without loss of generality, we assumed $U_n \in \mathbb{R}$. Note that the  $(\pm)$ pairs of minibands become decoupled if $\Omega  \rightarrow 0$, i.e.  in the limit of strong periodic potentials, $\left(C_n / |U_n|\right) \rightarrow 0$. As long as the energy separation between the miniband pairs dominates over the Zeeman splitting, i.e. $\sqrt{C_n^2 + |U_n|^2} \gg \Gamma/2$, we can treat $\widetilde{\Gamma}$ as the renormalized Zeeman splitting. 
From Eq. (\ref{RGam}) we see that $\widetilde{\Gamma} = \Gamma$ in the limit $\left(C_n / |U_n|\right) \rightarrow 0$, but, for finite $\left(C_n / |U_n|\right)$, the effective Zeeman splitting is renormalized to smaller values. To understand the physical  mechanism responsible for this behavior, we refer to Fig. \ref{FIG1}(c). 
Before the application of the periodic potential and magnetic field, the high energy states  are $\left|0,n,\uparrow\right>$ and $\left|0,-n,\downarrow\right>$. These two states cannot couple  directly  because $V(x)$ preserves spin and $\Gamma \sigma_x$ preserves the zone number, $n$, therefore the mixing between these states must rely on an indirect path involving both $\Gamma$ and $|U_n|$.  The periodic potential, $|U_n|$,  couples the (same-spin) upper and lower energy states, but   has to overcome an energy gap $2  C_n$. Hence, we expect a large effective Zeeman splitting $\widetilde{\Gamma}$ only if $\left(C_n /|U_n|\right) \lesssim 1$. 
Indeed, in the opposite limit, $\left(C_n / |U_n|\right) \gg 1$, we find a reduced effective Zeeman splitting, $\widetilde{\Gamma} \approx \left(|U_n| / C_n\right) \Gamma$. Since we are interested in realizing Majorana physics, which requires a Zeeman splitting $\widetilde{\Gamma} > \Delta$,  we focus of the strong periodic potential regime, $\left(C_n \ |U_n|\right) \lesssim 1$.

To investigate the emergence of topological superconductivity and Majorana bound states, we consider a  BdG Hamiltonian  with uniform induced pairing potential $\Delta$ and a normal component described by Eq. (\ref{Hn2}). Since a large effective Zeeman splitting is needed for the emergence of Majorana bound states, we focus on the regime $\left(C_n /|U_n|\right) \lesssim 1$, which implies $\widetilde{\Gamma} \gtrsim \Omega$.  This allows us to use quasi-degenerate perturbation theory \cite{Lowdin1951} to decouple the higher energy miniband pair from the lower energy pair. Note that we implicitly   incorporate the effects of the lower energy minibands on the higher energy pair, but a similar analysis can be done by explicitly keeping the lower energy pair in the effective model. To second order in $\Omega$, the effective Hamiltonian describing the higher energy minibands has the form
\begin{equation}
\begin{aligned}
H_{eff}^{\left(n\right)} =
& \left[\frac{\hbar^2 k^2}{2 m^*_{n+}} - \left( \widetilde{\mu}_{n+} - \frac{\Omega^2}{2 \sqrt{C_n^2 + |U_n|^2}} \right)      + \widetilde{\Gamma} \sigma_x\right] \tau_z \\ 
&+ \alpha_{n+} k \sigma_z  + \Delta \sigma_y  \tau_y. \label {Heff}
\end{aligned}
\end{equation}
This Hamiltonian corresponds to a simple, uniform Majorana nanowire model \cite{Lutchyn2010,Oreg2010} having effective parameters that are renormalized by the periodic potential [see Eqs. (\ref{ALP}-\ref{MUnT}), (\ref{RGam}), and (\ref{Omega})]. In the regime $\left(C_n/|U_n|\right) \ll 1$, the system undergoes a  topological phase transition at a (bare) critical Zeeman field
\begin{equation}
    {\Gamma}^2_c \approx 
    \left(\widetilde{\mu}_{n+}^2 + |\Delta|^2\right) 
    \left[1 + 
    \left(1 - \frac{\widetilde{\mu}_{n+}}{|U_n|}\right)
    \left(\frac{C_n}{|U_n|}\right)^2
    \right]. \label{GamC}
\end{equation}
Note that in the limit  $\left(C_n/|U_n|\right) \rightarrow 0$ we recover the ``standard'' expression of the critical field for a uniform Majorana nanowire \cite{Lutchyn2010,Oreg2010}. When ${\Gamma} > {\Gamma}_c$ the system is in a topological superconducting phase, with two zero-energy Majorana bound states localized at the edges.
 
\begin{figure}[t]
\begin{center}
\includegraphics[width=0.48\textwidth]{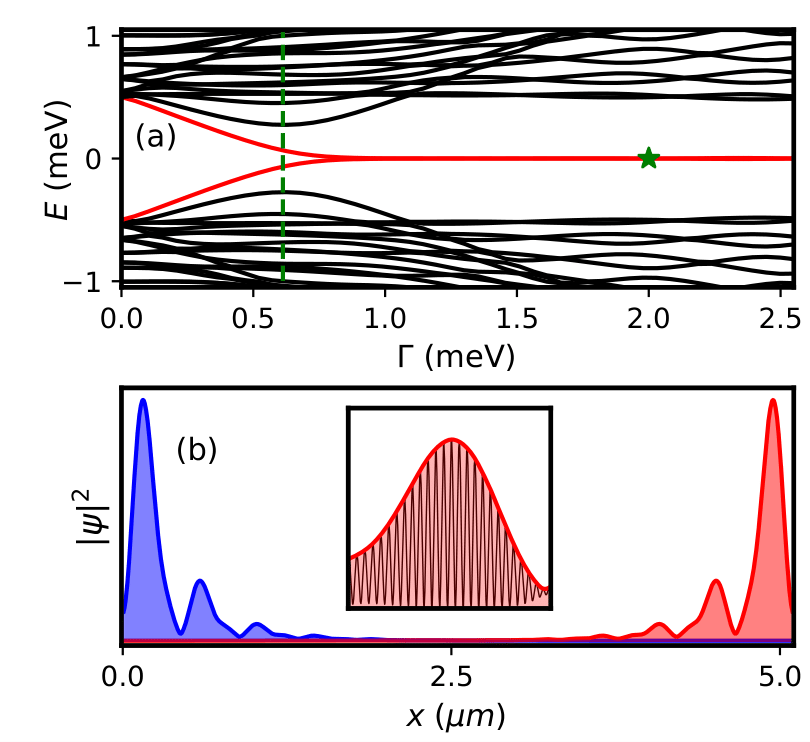}
\end{center}
\vspace{-5mm}
\caption{(a) Low-energy spectrum of a finite system described by Eq. (\ref{HamBdG}) as a function of the applied (bare) Zeeman splitting $\Gamma$. The system parameters are $m^* = 0.026~ m_o$, $\alpha = 20$ meV$\cdot$nm, $\Delta = 0.5$ meV, $\ell = 25$ nm, $L = 5.1~ \mu$m, and $C_1/U_1 = 0.5$. The chemical potential is $\mu = 103.5$ meV, corresponding to the bottom of the $|n|=1$ higher energy miniband pair, i.e. $\widetilde{\mu}_{1+}=0$. The green dashed line shows the critical Zeeman splitting predicted by Eq. (\ref{GamC}), which coincides with the minimum of the bulk gap, as expected. (b) Envelopes of the MBS wave functions corresponding to the green star in (a). The inset shows a zoom-in of the second maximum of the right Majorana. Note the highly oscillatory nature of the Majorana wave function  (black lines).}
\label{FIG2}
\vspace{-3mm}
\end{figure}

To illustrate the emergence of Majorana bound states in a high energy miniband, we solve numerically the BdG problem described by Eq. (\ref{HamBdG}) for a finite system using the finite difference method. For simplicity and clarity we first consider an idealized periodic potential composed of a single harmonic of the form $V\left(x\right) = 2U \cos\left(\pi x / \ell\right)$, with $\ell = 25$ nm, so that $\widetilde{V}_n = U \delta_{n,\pm1}$.  
Other system parameters are $m^* = 0.026~ m_o$, $\alpha = 20$ meV$\cdot$nm, $\Delta = 0.5$ meV,  $L = 5.1~ \mu$m, $C_1/U_1 = 0.5$. The chemical potential is $\mu = 103.5$ meV, which corresponds to $\widetilde{\mu}_{1+}=0$, i.e. the chemical potential is set to the bottom of the $|n|=1$ higher energy miniband pair. The results are shown  in Fig. \ref{FIG2}.
Upon applying a sufficiently high Zeeman field, a pair of Majorana modes emerges at zero energy [see Fig. \ref{FIG2}(a)]. Note that Majorana bound states could not emerge at such a high chemical potential ($\mu = 103.5$ meV) in a uniform system, as the required Zeeman field would completely destroy superconductivity in the parent superconductor. Rather, the periodic potential has expanded the parameter space consistent with topological superconductivity, as pointed out in Refs. \cite{Adagideli2014,Levine2017,Escribano2019}. The green dashed line in Fig. \ref{FIG2}(a) shows the critical Zeeman splitting predicted by the analytical expression in Eq. (\ref{GamC}). Note that this value coincides with the minimum of the bulk gap, which occurs at $\Gamma > \Delta$ due the additional $\left(C_n/|U_n|\right)$ contribution in Eq. (\ref{GamC}). The wave functions of the two MZMs corresponding to $\Gamma = 2~$meV, which are obtained using the Majorana representation \cite{Moore2018} of the lowest energy states, are shown in Fig. \ref{FIG2}(b). Note that the broad blue and red curves, which  represent the modulus squared of the envelope functions corresponding to the two MZMs localized at the ends of the wire,  
are very similar to Majorana wave functions emerging in a uniform system. However, a zoom-in of the second maximum of the right (red) Majorana mode [see the inset of Fig. \ref{FIG2}(b)] reveals the highly oscillatory nature of the wave function, which oscillates with a wavelength $\ell$. This rapidly oscillating  nature, which is indicative of large $k$ components, represents the source of the enhanced (renormalized) spin-orbit coupling in Eq. (\ref{ALP}).  

\begin{figure}[t]
\begin{center}
\includegraphics[width=0.48\textwidth]{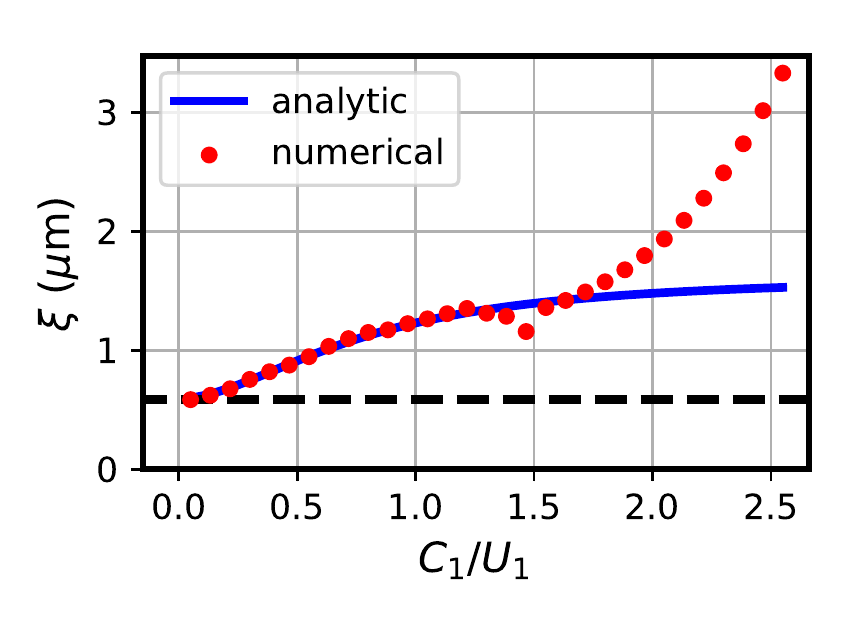}
\end{center}
\vspace{-5mm}
\caption{Localization length $\xi_{1+}$ as a function of $C_1/U_1$ for a system with $\Gamma = 2$ meV and $\widetilde{\mu}_{1+} = 0$. The red dots show the analytical result given by Eq.  (\ref{XI}), while the blue line corresponds to the numerical solution extracted by fitting the envelope of the Majorana wavefuction, $|\psi| \propto e^{-x/\xi_{1+}}$. The system parameters are the same as in Fig. \ref{FIG2}, except $L = 20~\mu$m and $C_1/U_1$ varies. For comparison, the localization length of a uniform wire with the same \textit{bare} parameters ($m^*$, $\alpha$, and $\Delta$) and $\mu=0$ is shown as a black dashed line.}
\label{FIG3}
\vspace{-3mm}
\end{figure}

\subsection{Majorana localization length} 

A key parameter that characterizes the Majorana bound states is the localization length $\xi$ representing the characteristic length scale of the MBS. This localization length controls, among other things,  the amplitude of the energy splitting oscillations due to the partial overlap of the MBSs localized at the opposite ends of a finite wire. 
A natural question is  how does the localization length of a periodic Majorana structure compare with the localization length of the corresponding uniform system? On the one hand, the reduced effective mass [see Eq. (\ref{MEFF})] favors delocalization, while, on the other hand, the increased spin-orbit coupling [see Eq. (\ref{ALP})] enhances the Majorana localization. 
 To determine the relative role of these effects, we study (analytically and numerically) the solutions of the effective Hamiltonian (\ref{Heff}). Details can be found in Appendix \ref{LOCAL}. We find the the localization length of the MBSs associated with the higher energy miniband pair ($n+$) has the form
\begin{equation}
    \xi_{n+} \sim \ell_{SO} \left(\frac{\Gamma}{\Delta}\right)
    \sqrt{
    1 + 
    \frac{4 \widetilde{\mu}_{n+} \widetilde{E}_{SO,n+}}{\Gamma^2} + 
    \frac{4 \widetilde{E}_{SO,n+}^2}{\Gamma^2}
    },               \label{XI}
\end{equation}
where $\ell_{SO} = \hbar^2 / \left(m^* \alpha\right)$ is the \textit{bare} spin-orbit length, while $\widetilde{\mu}_n+$ and $\widetilde{E}_{SO,n+}$ are the \textit{renormalized} chemical potential (\ref{MUnT}) and spin-orbit energy (\ref{ESO}), respectively.  Note that a similar calculation can be done for $\xi_{n-}$. The first two factors in Eq. (\ref{XI}) (i.e., those outside the square root sign) are \textit{bare} parameters entering the original BdG Hamiltonian (\ref{HamBdG}).  The last factor (i.e., the square root) contains \textit{renormalized} parameters and leads to a moderate increase of the localization length. 
 We verify Eq. (\ref{XI})  numerically by fitting the envelope of the Majorana wave function to an exponential, $|\psi|^2 \propto e^{-2x/\xi}$. The results are shown in Fig. \ref{FIG3} as a function of $(C_1/U_1)$ for fixed Zeeman field, $\Gamma = 2$ meV, and $\widetilde{\mu}_{1+} = 0$. Note the  excellent agreement between the analytical and numerical results for $(C_1 / U_1) \lesssim 1.5$, which corresponds to the strong periodic potential regime. Above this threshold, the localization length increases strongly as the critical Zeeman field $\Gamma_c$ approaches $\Gamma=2~$meV. Note that the dashed line shows the localization length of a uniform wire [$V(x)=0$, $\mu = 0$] with the same bare parameters as the periodic system. 
In the limit $C_1/U_1 \rightarrow 0$, the periodic and uniform systems have the same localization length, $\xi \approx 600$ nm, while the localization length of the periodic system increases with increasing $C_1/U_1$. At $C_1/U_1 = 1$, $\xi$ is roughly double for the periodic system as compared to the uniform wire. However, this is a rather moderate increase, particularly considering that the renormalized effective mass $m^*_{1+} = 0.0016~ m_o$ is significantly smaller than the bare effective mass,  $m^* = 0.026~ m_o$. Finally, we note that, while the results shown in Figs. (\ref{FIG2}) and (\ref{FIG3}) are based on an idealized periodic potential of the form $V\left(x\right) = 2U \cos\left(\pi x/\ell\right)$, the basic physics discussed above holds for generic periodic potentials. 

\begin{figure}[t]
\begin{center}
\includegraphics[width=0.48\textwidth]{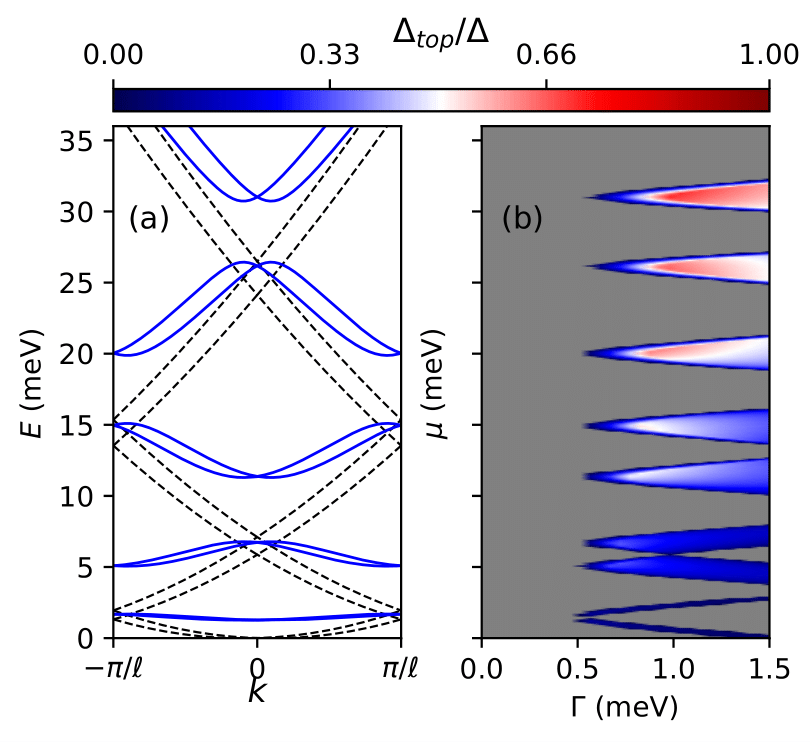}
\end{center}
\vspace{-5mm}
\caption{(a) Low energy spectrum of a normal wire in the absence (black dashed lines) and presence (blue solid lines) of a periodic potential. (b) Topological phase diagram of the periodic system as a function of chemical potential $\mu$ and Zeeman field $\Gamma$. Gray regions are topologically trivial. The color of the topological regions is determined by the size of the topological gap. The system parameters are $m^* = 0.023~m_0$, $\alpha = 10~\text{meV}\cdot\text{nm}$, $\Delta = 0.5~\text{meV}$, $\ell = 100~\text{nm}$, $L_{bar} = 15~\text{nm}$, and $V_o = 20~\text{meV}$.}
\label{FIG4}
\vspace{-3mm}
\end{figure}

\subsection{Topological phase diagram in systems with rectangular periodic potential}

To investigate the effect of the periodic potential on the topological phase diagram, we consider a system with rectangular periodic potential, as  shown in Fig. \ref{FIG1}(a). Explicitly, we have 
\begin{equation}
    V\left(x\right) = 
    \left.
  \begin{cases}
    V_o, &  0 \leq x \leq L_{bar} \\
    0, & L_{bar} < x < \ell 
  \end{cases}
  \right. , \label{POT}
\end{equation}
where $V_o$ and $L_{bar}$ are the height and length of each potential barrier, respectively, and $ V\left(x+ \ell\right)=V\left(x\right)$. The Fourier components of the potential are 
\begin{equation}
    \widetilde{V}_n = \frac{-i V_o}{2 \pi n} 
    \left[ 
    \exp \left( \frac{i 2 \pi n L_{bar}}{\ell}\right)  - 1
    \right ].
\end{equation}
The energy spectra of the normal system  with and without the periodic potential are shown in Fig. \ref{FIG4}(a)  as blue solid lines and black dashed lines, respectively. In the absence of the periodic potential, the dispersion is quadratic and there are no energy gaps. Applying a periodic potential opens energy gaps at the zone center and at the zone boundaries inducing (pairs of) minibands. 
\begin{figure}[t]
\begin{center}
\includegraphics[width=0.48\textwidth]{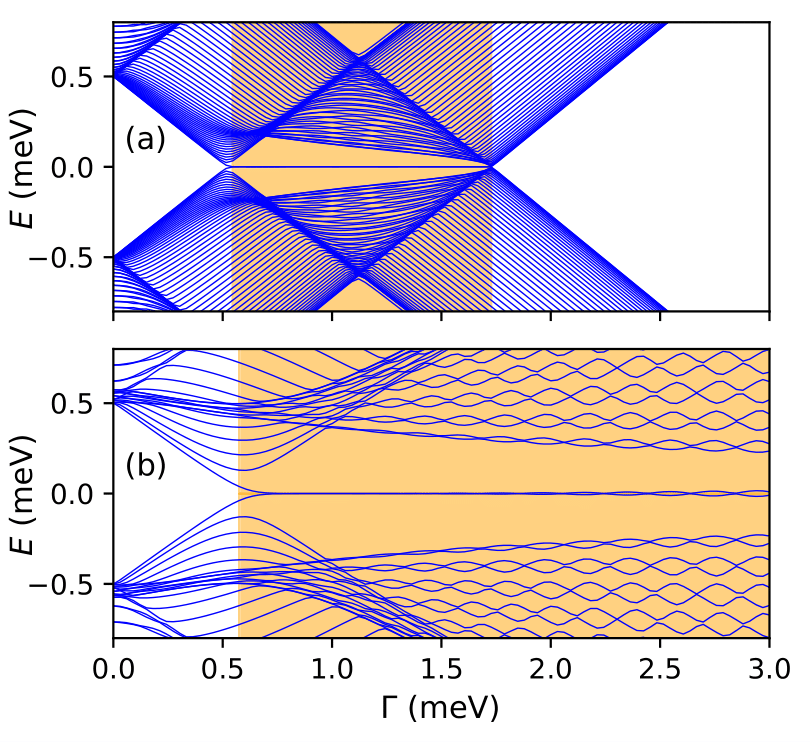}
\end{center}
\vspace{-5mm}
\caption{Energy spectrum of a finite system of length $L = 6~\mu$m as function of the Zeeman field for a chemical potential (a) $\mu = 6.73$ meV (corresponding to the top of the second pair of minibands) and (b) $\mu = 31.02$ meV (bottom of the fifth miniband pair). The topological phase (corresponding to the orange shading) supports a pair of zero-energy Majorana modes. Note that the topological gap is significantly larger in panel (b), i.e., in the topological phase associated with the fifth miniband [see also Fig. \ref{FIG4}(b)].}
\label{FIG5}
\vspace{-3mm}
\end{figure}
As noted above, the dispersion of the minibands is similar to that of uniform Rashba nanowires, except that the effective parameters are renormalized and miniband-dependent [see Eqs. (\ref{NRG}-\ref{MUnT})]. Each of the minibands can support Majorana bound states, as long as the chemical potential is close to its bottom/top and the Zeeman splitting is strong-enough. For a given set of parameters, we use the Chern-Simon invariant \cite{Kitaev2001,Chiu2016}, $\mathcal{Q}$, to determine whether the system is topologically trivial ($\mathcal{Q} = 1$) or non-trivial ($\mathcal{Q} = -1$). Fig. \ref{FIG4}(b) shows the calculated phase diagram as a function of Zeeman splitting, $\Gamma$, and chemical potential, $\mu$. Topologically trivial regions are shown in gray, while the topologically non-trivial regions are colored using a color-scale that indicates the size of the topological gap. Several features are worth pointing out. 
First, we emphasize that the emergence of a  low-field topological phase is  associated with the chemical potential being near one of the  miniband edges, i.e., either the one at $k = 0$ or the  $k = \pm \pi / \ell$ band edge. The critical Zeeman field $\Gamma_c(\mu)$ has minima near each of these band edges. Note that the corresponding uniform system only supports a  low-field topological phase for low values of $\mu$, i.e., for chemical potential values near the bottom of the conduction band. 
Second, the areas of the parameter space that support topological superconductivity typically increase as one reaches higher energy minibands. This is due to the increasing of the miniband width with energy.
For example, the lowest energy miniband has a bandwidth of less than $0.5~\text{meV}$. For any (large-enough) value of the Zeeman field, the single miniband occupancy condition (consistent with the emergence of topological superconductivity) is only satisfied within a narrow chemical potential window comparable to the bandwidth.  
The increase of the topological phase with the miniband index is further  illustrated in Fig. \ref{FIG5}, which shows the dependence of the low-energy spectrum of a finite wire on the applied Zeeman field for two different values of the chemical potential corresponding  to the $k=0$ band edge of the (a) second and (b) fifth pair of minibands, respectively. 
Both sets of parameters support topological superconductivity and zero-energy MBSs, but in the case of the second miniband [panel (a)] this occurs over a rather narrow range of Zeeman fields. Indeed,  the system in panel (a) becomes non-topological for $\Gamma > 1.7~\text{meV}$, when the $k = \pm \pi / \ell$ band edge crosses the Fermi level. 
The third important feature of Fig. \ref{FIG4}(c) is the increasing of the topological gap as the chemical potential moves into the higher energy minibands. The larger topological gap is a consequence of the larger spin-orbit energy characterizing the higher energy minibands [see Eq. (\ref{ESO})]. We conclude that the presence of a periodic potential (i) expands the (low-field) parameter region that supports a topological superconducting phase and (ii) can enhance the size of the (low-field) topological gap. Note that throughout this work we are particularly interested in the low-field regime ($\Gamma \lesssim 5\Delta$), as strong magnetic fields are detrimental to superconductivity, can lead to the complete collapse of the parent superconducting gap, and, therefore, are less relevant to understanding the physics of experimentally realizable structures.  

\begin{figure}[t]
\begin{center}
\includegraphics[width=0.48\textwidth]{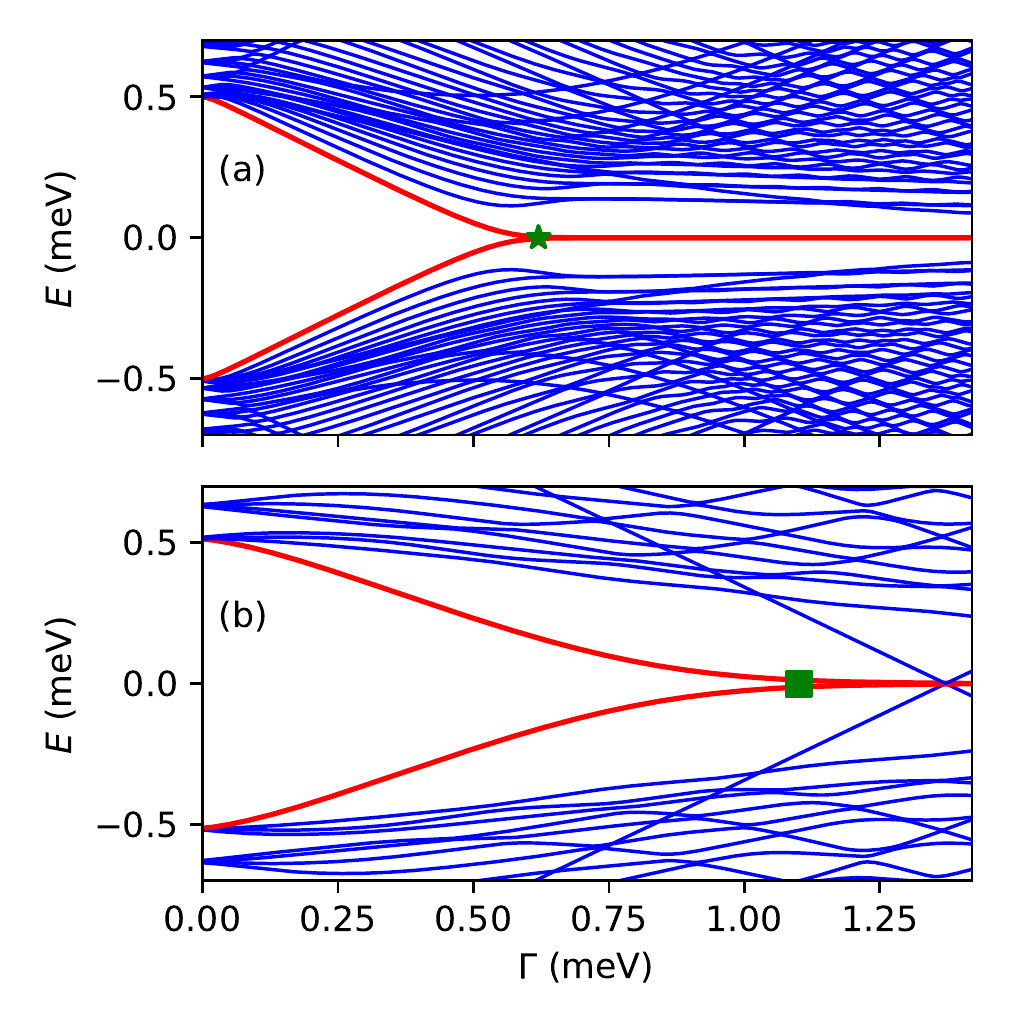}
\end{center}
\vspace{-5mm}
\caption{(a) Low-energy spectrum as a function of the applied Zeeman field for a system with soft confinement (and no periodic potential). (b) Low-energy spectrum of a periodic system with soft confinement.  The periodic potential has the form $V\left(x\right) = V_o \cos(\pi x / \ell)$, with $V_o = 8$ meV and $\ell = 20$ nm, while the soft confinement has a slope $\kappa = 2~\text{meV}/\mu\text{m}$. Both systems support  ABS modes that collapse toward (and stick to) zero-energy as the Zeeman splitting increases. Note, however, that  in the periodic system the ABS collapses to zero-energy at a significantly larger Zeemen field.}
\label{FIG6}
\vspace{-3mm}
\end{figure}

\subsection{Soft confinement and partially-separated Andreev bound states in periodic systems} \label{Soft}

Systems with inhomogeneous parameters are known to give rise to low-energy Andreev bound states (ABSs) in the topologically-trivial phase \cite{Kells2012,Bagrets2012,Liu2012,Roy2013,Adagideli2014,Moore2016,Moore2018,Stanescu2018b,Fleckenstein2018,Pan2019,Vuik2019}. Of particular interest are the trivial near-zero energy modes that mimic the local phenomenology of Majorana bound states, the so-called  partially-separated Andreev bound states (ps-ABS) or quasi-Majorana modes \cite{Stanescu2018b,Vuik2019}. In the Majorana representation, these low-energy states are characterized by Majorana components that are partially separated in space, unlike ``standard'' ABSs, which consist of highly overlapping Majorana components. The energy splitting of the low-energy ABSs (including the ps-ABSs) is sensitive to local perturbations, indicating that these states do not share the topological protection of well separated MBSs. Here, we study the emergence of low-energy ps-ABSs  in a periodic system due to soft confinement and compare the properties of these states with those of ps-ABSs  emerging in ``conventional'' nanowires with soft confinement and no periodic potential. For clarity, we consider an idealized soft confinement given by the potential
\begin{equation}
    V^\prime \left(x\right) = 
    - \kappa x,    \label{Vprime}
\end{equation}
where $\kappa$ is the slope of the potential.
The potential given by Eq. (\ref{Vprime})  has the property that, for sufficiently long wires, induced  low-energy states are independent of  the chemical potential, up to an overall spatial shift.
For small-enough values of $\kappa$ we expect nearly zero energy ps-ABSs emerging for values of the Zeeman splitting slightly above $\Delta$, since there will be a sufficiently-wide region that \textit{locally} satisfies the topological condition $\Gamma > \sqrt{[\mu-V(x)]^2 + \Delta^2}$. Increasing $\kappa$ shrinks the region of space where the topological condition is met (for a given value of $\Gamma$) and the ps-ABS collapses toward zero energy at larger values of the Zeeman field. 

The low-energy spectra of a system with soft confinement in the (a) absence and (b) presence of the periodic potential are compared in Fig. \ref{FIG6}. 
The periodic potential has the form $V\left(x\right) = V_o \cos(\pi x / \ell)$, with $V_o = 8$ meV and $\ell = 20$ nm.
 We notice that the low-energy spectrum in Fig. \ref{FIG6}(a), i.e., in the absence of a periodic potential, has a zero energy state emerging at a Zeeman field  just above the induced pairing $\Delta=0.5~$meV. By contrast, the periodic system supports a near-zero energy state only above $\Gamma \approx 1.1~$ meV. This behavior suggests that the presence of a periodic potential reduces the low-field parameter region that supports ps-ABSs (i.e. quasi-Majoranas). 

\begin{figure}[t]
\begin{center}
\includegraphics[width=0.48\textwidth]{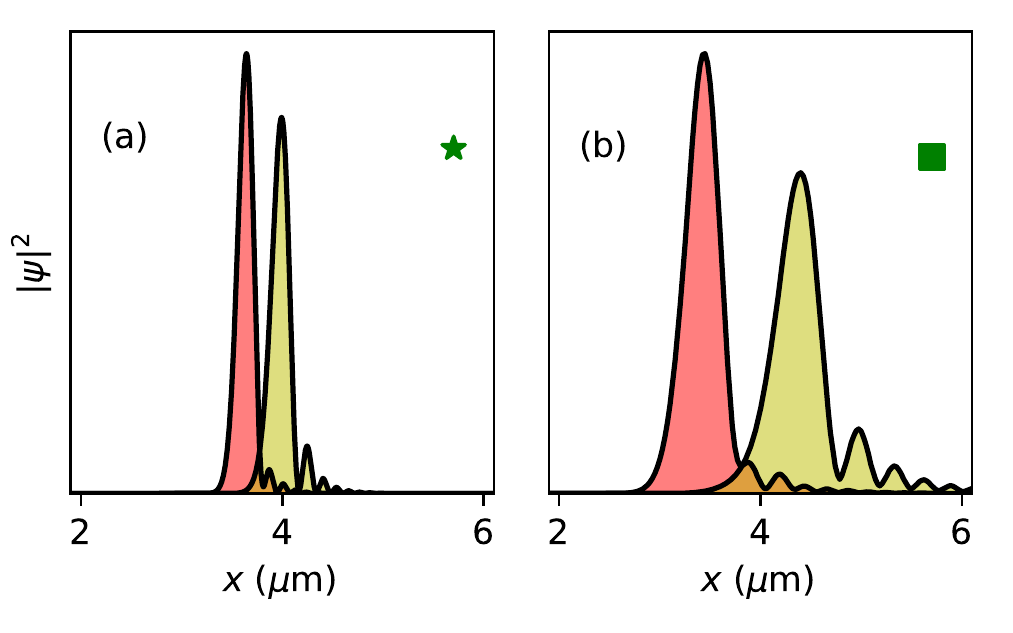}
\end{center}
\vspace{-5mm}
\caption{(a) Majorana wave functions corresponding to the low-energy ps-ABS marked by a green star in Fig. \ref{FIG6}(a).  (b) Envelopes of the Majorana wave functions corresponding to the low-energy ps-ABS marked by a green square in Fig. \ref{FIG6}(b). Note that the actual wave function oscillates rapidly, as illustrated in Fig. \ref{FIG2}(b). The main Majorana peaks are wider in the periodic system [panel (b)] as a result of a smaller effective mass. In turn, this requires a higher Zeeman field for satisfying the partial separation condition associated with the collapse of the ps-ABS to zero energy \cite{Stanescu2018b} (see Fig. \ref{FIG6}). }
\label{FIG7}
\vspace{-3mm}
\end{figure}

To understand this behavior, we calculate the Majorana components of the ps-ABSs marked by the green star and square in Fig. \ref{FIG6} (a) and (b), respectively. The results are shown in  Fig. \ref{FIG7}. 
Both states are characterized by a partial separation of the Majorana components larger than the widths of the corresponding main peaks of the Majorana wave functions, which is a necessary condition for the collapse of the ps-ABS to zero energy \cite{Stanescu2018b}. 
The key difference between the states in Fig. \ref{FIG7}(a) and and those in Fig. \ref{FIG7}(b) is that the ``conventional'' ps-ABS is characterized by Majorana peak widths that are significantly narrower than the corresponding peaks of the ps-ABS emerging in the periodic system. Consequently, in a periodic wire with soft confinement the partial separation condition \cite{Stanescu2018b} is realized at higher values of the Zeeman field,  as compared to a ``conventional'' system, as explicitly shown in Fig. \ref{FIG6}. Finally, we note that width of the (main) Majorana peak, which controls the Zeeman field associated to the collapse to zero energy of the ps-ABS, is determined by the effective mass. In a periodic system, the effective mass can be significantly reduced, particularly for higher energy minibands [see eq. (\ref{MEFF})].   Consequently, the presence of a periodic potential reduces the probability of (accidental) ps-ABSs emerging in the (topologically-trivial) low-field regime, and, implicitly, reduces the likelihood of getting false positives when searching for Majorana zero modes. Quantitative estimates of this superlattice-induced reduction require a more realistic modeling of the hybrid system. 

\subsection{Robustness against disorder} \label{DIS_1D}

Our next objective is to investigate the robustness against disorder of the topological superconducting phase realized in a periodic system and compare it with the robustness of the corresponding phase emerging in uniform structures. We note that the presence of disorder, which, to some degree, is inevitable in real systems, can lead to the reduction of the topological gap \cite{Sau2012,Hui2015} and the emergence of trivial low-energy modes \cite{Brouwer2011,Bagrets2012,DeGottardi2013,Pan2019b}. To investigate the effects of disorder,  we consider a correlated Gaussian disorder potential, $V^\prime\left(x\right)$, characterized by 
\begin{eqnarray}
    \left< V^\prime \left(x_i\right) \right > &=& 0, \\
    \left< V^\prime \left(x_i\right) V^\prime \left(x_j\right) \right > &=& 
    U^2 \exp \left(-\frac{\left| x_i - x_j^\prime\right|}{L_{dis}}\right),
\end{eqnarray}
where $U$ is the disorder strength, $x_i$ is the position along the wire corresponding to lattice site $i$, and $L_{dis}$ is the disorder correlation length scale.  In our numerical calculations we use $L_{dis} = 50~\text{nm}$, which is a length scale comparable to the typical diameter of a semiconductor nanowire. The correlated disorder is numerically implemented using the scheme described in Ref. \cite{Deserno2002}.  The other system parameters are the same as in Fig. \ref{FIG4} and the wire length is $L = 5~\mu\text{m}$. For comparison, we also consider a ``conventional'' disordered wire having the same system parameters and disorder potential, but no periodic potential. 

\begin{figure*}[t]
\begin{center}
\includegraphics[width=1.\textwidth]{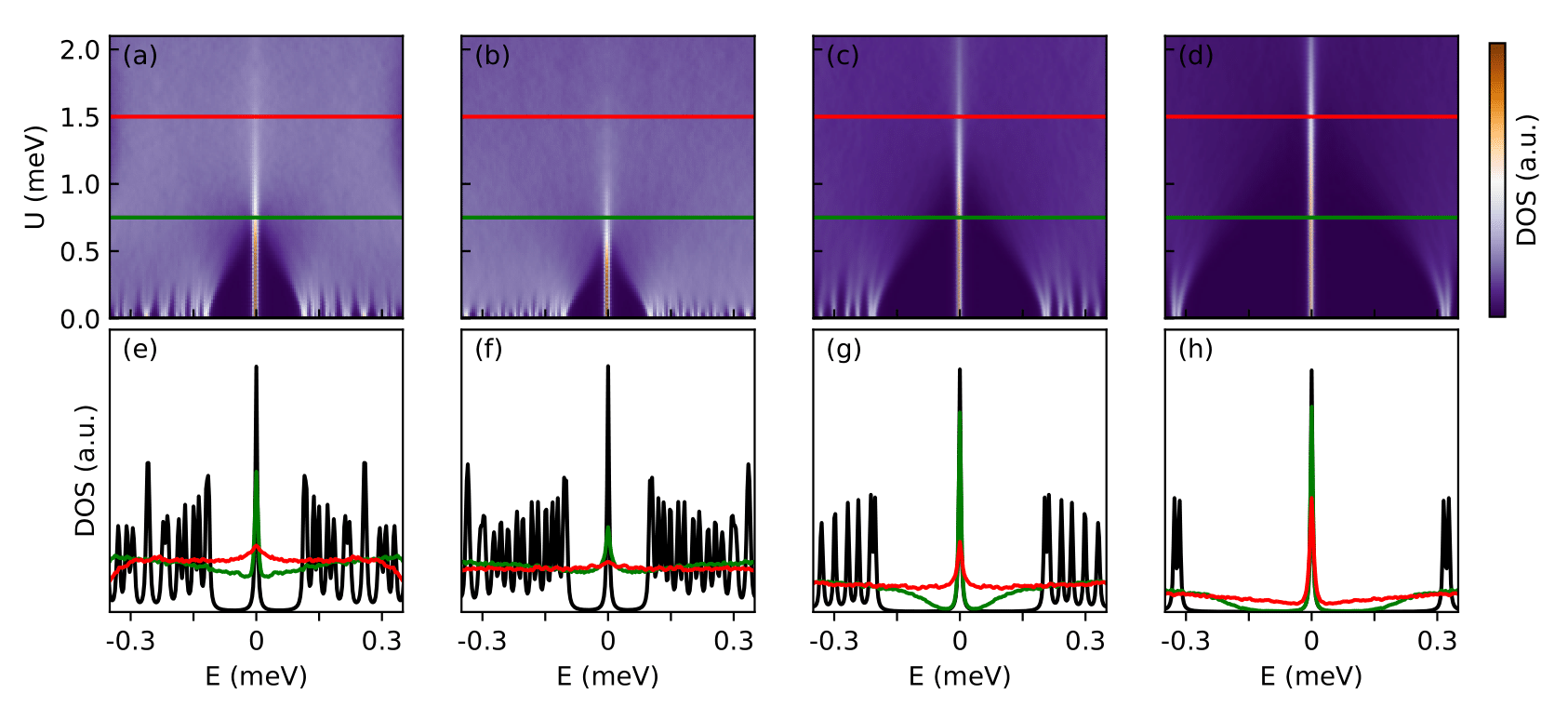}
\end{center}
\vspace{-5mm}
\caption{Density of states averaged over 1000 disorder realizations as function of the disorder strength $U$ for (a) a uniform system with chemical potential $\mu=0$ and (b-d) a periodic structure with chemical potential set to the second (b), third (c), and fourth (d) $k=0$ miniband edges. The 
 Zeeman splitting is fixed at $\Gamma = 1.2~\text{meV}$, the length of the wire is $L = 5~\mu\text{m}$, the disorder correlation length is $L_{dis} = 50~\text{nm}$, and other system parameters are the same as Fig. \ref{FIG4}.  The bottom row (e-h) shows line cuts corresponding to the colored lines in the top row: $U=0$ (black lines), $U=0.75~$meV (green), and $U=1.5~$meV (red).
Note that the topological gap and the Majorana zero-energy peak corresponding to higher-energy minibands of the periodic system [panels (c,d) and (g,h)] are significantly more robust against disorder than their counterparts emerging in the uniform system [panels (a) and (e)].}
\label{FIG8}
\vspace{-3mm}
\end{figure*}

First, we calculate the density of states (DOS) averaged over disorder (using 1000 realizations) as a function of the disorder strength for a fixed value of the Zeeman field, $\Gamma = 1.2~\text{meV}$, well inside the topological phase of a clean system.  The results are shown in 
Fig. \ref{FIG8}  for (a) a uniform system with  chemical potential $\mu = 0$ and (b-d) a periodic systems  with chemical potential corresponding to the bottom of the second, third, and fourth $k=0$ miniband edges, respectively. 
In the absence of disorder ($U=0$), the system is in the topological phase and supports a pair of zero-energy MBSs localized at the edges of the finite wire.
The presence of the Majorana modes is signaled by a sharp zero-energy peak in the density of states (see Fig. \ref{FIG8}). 
We also note the larger topological gap associated with the higher energy minibands of the periodic system, which is due to a larger effective spin-orbit coupling. Indeed, the zero-disorder topological gap in the third and fourth $k=0$ minibands [panels (c) and (d) in Fig. \ref{FIG8}] is significantly larger than the corresponding gap in the uniform system [panel (a)]. As the disorder is turned on, the topological gap collapses as low-energy bound states (localized by the disorder potential) start to populate the gap. Note that the collapse of the gap to zero energy occurs at significantly higher disorder strengths in the periodic system (with chemical potential inside higher-energy minibands) as compared to the uniform case. This is consistent with the corresponding size of the topological gap at $U=0$. 
 For clarity,  in the bottom row of Fig. \ref{FIG8} we provide line cuts of the (disorder-averaged) density of states at fixed disorder strengths.  One clearly notices the zero energy peak  associated with the presence of MBSs and the collapse of the topological gap in the presence of disorder. Remarkably, in the periodic system with a chemical potential inside the fourth miniband [panel (h)], a finite topological gap survives at $U=0.75~\text{meV} >\Delta$ and the zero-energy peak is well defined even at $U=1.5~$meV. We also note that the (average) density of states corresponding to higher energy minibands is reduced relative to the uniform system DOS due to a smaller (renormalized) effective mass in the periodic system. This results in a large ratio between the zero-energy peak height and the background density of states corresponding to high-energy minibands of the periodic system [see, e.g, Fig. \ref{FIG8}(h)].
 
The results discussed above suggest that  a periodic system can support (low-field) topological superconductivity and Majorana zero modes that are significantly more robust against disorder than their counterparts realized in a uniform system (having the same parameters), provided the chemical potential is tuned into the higher-energy minibands.
An interesting problem that cannot be settled based on our density of states analysis concerns the survival of the zero-energy peaks above the disorder strength corresponding to the collapse of the topological gap (see Fig. \ref{FIG8}). While it is tempting to attribute the peak to MBSs localized at the ends of the wire, which is definitely the case for weak disorder, one has to keep in mind that generic class D systems are known to have zero-energy peaks in their density of states even in the absence of topological MBS localized at the boundaries. This phenomenon can be understood in terms of the Griffiths effect \cite{Motrunich2001,Sau2012}, i.e. the disorder potential causing fluctuations in the chemical potential that generate short topologically trivial and non-trivial regions throughout the wire.  Highly overlapping low-energy MBS emerge at domain walls between these regions. 
It has been shown that a power-law peak in the density of states at $E=0$ is expected to occur due to such disorder induced states, in contrast with the sharp peak associated with topological MBSs localized at the edges of the wire \cite{Motrunich2001}. 

\begin{figure*}[t]
\begin{center}
\includegraphics[width=1.\textwidth]{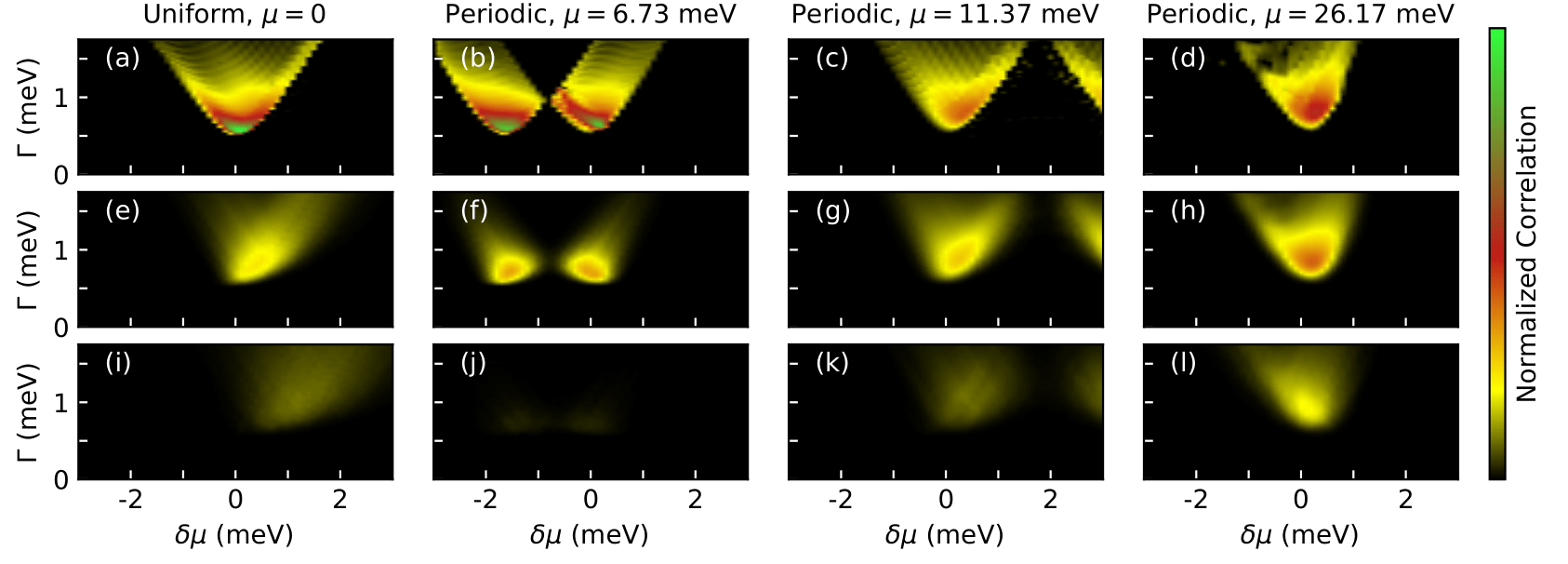}
\end{center}
\vspace{-5mm}
\caption{\textit{Correlation phase diagrams} for uniform and periodic systems as a function of chemical potential and Zeeman splitting. The correlation strength $C$ is defined by Eq. (\ref{Correlation}) and is averaged over 300 disorder realizations, with each panel being normalized to the maximum value corresponding to a clean, uniform system -- panel (a). The disorder strength is $U = 0$ for the panels in the top row (a-d),  $U=0.5~\text{meV}$ (middle row), and $U=1.0~\text{meV}$  (bottom row). Note that the edge-to-edge correlations are suppressed most rapidly in the periodic system with low chemical potential (second column), followed by the uniform system (first column). In the periodic system with higher-energy minibands (third and fourth columns) the correlations decrease with disorder at a lower rate,  indicating an increases robustness of MBSs against disorder.}
\label{FIG9}
\vspace{-3mm}
\end{figure*}

To clearly disentangle the contributions generated by topological MZMs from those associated with local MBS pairs (including ps-ABSs, or quasi-Majoranas), we introduce the following  edge-to-edge correlation function
\begin{equation}
    C =~  \max\left(0,\widetilde{C}\right), \label{Correlation}
\end{equation}
with
\begin{equation}
\begin{aligned}
    \widetilde{C} =~ & 
    |\psi_{1,L}|^2 |\psi_{1,R}|^2 f(E_1,\omega_o,\Omega) \\
    -& \sum_\lambda^{\lambda \neq 1} 
    |\psi_{\lambda,L}|^2 |\psi_{\lambda,R}|^2 f(E_\lambda,\omega_o,\Omega),  \label{Corr}
\end{aligned}
\end{equation}
where $\psi_{\lambda,L(R)}$ is the left (right) edge component of the  positive energy state $\lambda$ and $f$ is a weight function (explicitly defined below) peaked at zero-energy and characterized by an energy window $\omega_0$. Note that $C\neq 0$ signals that the lowest energy mode has nonzero weight at both ends of the wire, i.e. it consists of well-separated MBSs. Possible non-local correlations associated with higher energy states result in $\widetilde{C} < 0$ [see eq. (\ref{Corr})], hence $C=0$. 
Also note that the lowest energy state can have a finite energy $\epsilon$ and still generate a well-defined  edge-to-edge correlation, as long as $\epsilon <\omega_0$. This property is extremely useful for disentangling well-separated MBSs and ps-ABSs in finite (relatively short) wires, as both types of states may be characterized by finite energy splitting oscillations. 
The weight function is defined as
\begin{equation}
    f\left(E,\omega_o,\Omega\right) = 
    H(E,-\omega_o,\Omega) - H(E,\omega_o,\Omega),
\end{equation}
with 
\begin{equation}
    H\left(E,\omega_o,\Omega\right) = 
    \left(
    1 + \exp{\left[-2\left(E-\omega_o\right)/\Omega\right]}
    \right)^{-1}. \label{Heavi}
\end{equation}
Note that Eq. (\ref{Heavi}) becomes the Heaviside step function in the limit $\Omega \rightarrow 0$. In the numerical calculations we use $\omega_o = 2~ \text{meV}$ to define the range of relevant low-energy states and $\Omega = 0.2 ~\text{meV}$ to define the smooth transition region. Also, we define $\psi_{\lambda,L(R)}$ as the total weight (including summations over the spin and particle-hole degrees of freedom) within the leftmost (rightmost) $200~\text{nm}$ segment of the wire. 
  
Calculated edge-to-edge correlations averaged over 300  disorder realizations are shown in Fig. \ref{FIG9} for both the periodic and uniform systems. The top panels correspond to a clean system ($U=0$), while the middle and bottom panels correspond to $U=0.5~$meV and $U=1~$meV, respectively. The length of the wire is  $L = 2~\mu\text{m}$, while other system parameters are the same as in Fig. \ref{FIG8}.  First, we note that for a clean system (see top panels in Fig. \ref{FIG9}) the areas with $C\neq 0$ correspond to the the topological superconducting phase in Fig. \ref{FIG4}(b), demonstrating that the edge-to-edge correlations are generated by Majorana modes localized at the ends of the wire. To emphasize this property, we will use the term \textit{correlation phase diagram} to designate the map $C(\mu, \Gamma)$.
Note that the large correlations corresponding to the green regions in panels (a) and (b) are due to strongly localized  MBSs characterizing the corresponding clean systems. By comparison,  the (clean)  MBSs associated with higher energy minibands [panels (c) and (d)] have larger localization lengths, hence lower values of $|\psi_{1,L(R)}|^2$ and, implicitly, $C$. 
Upon introducing disorder, the edge-to-edge correlation is reduced in all cases, essentially due to the hybridization of end-of-wire MBSs with disorder-induced low-energy states. In addition, one can notice (primarily in the uniform system) a shift of the correlated area toward  larger values of $\mu$, which is consistent with the findings of Ref. \cite{Adagideli2014}.
However, the most  important feature is that the suppression of $C$ does not occur at the same rate for all parameter regimes. While in a periodic system with low chemical potential (second column in Fig. \ref{FIG9}) the correlations are suppressed even faster than in the uniform system (first column), periodic systems with a chemical potential within higher energy minibands (columns three and four) are characterized by correlations that decrease at a lower rate than the uniform system, which signals the increased robustness of MBSs against disorder.
Note that while panels (i) and (k) have similar correlation levels at $U = 1~\text{meV}$, the periodic system (third column) is characterized by correlations decreasing with the disorder strength at a significantly slower rate than the correlations in the homogeneous system (first column).  
We  conclude that topological phases and Majorana zero modes realized in periodic systems can be more robust against disorder than their counterparts emerging in uniform system, provided the chemical potential lies within a sufficiently high energy miniband. 

\section{Majorana Waveguides} \label{Waveguide} 

Our detailed analysis of the periodic Majorana system based on the 1D toy model given by Eqs. (\ref{HamBdG}) and (\ref{Ho}) has demonstrated that the presence of a periodic potential could provide significant advantages for realizing Majorana zero modes in semiconductor-superconductor structures, such as  increasing the low-field parameter range consistent with topological superconductivity,  enhancing the topological gap and the robustness of the Majorana modes against disorder, and reducing the (low-field) parameter space associated with the emergence of topologically trivial low-energy states.  However, our analytical expressions for the renormalized effective parameters, corroborated by the numerical results,  indicate that these potential benefits can only be obtained if (i) the applied periodic potential is strong-enough and (ii) the chemical potential is tuned within one of the higher energy minibands. In practice, realizing the first condition is highly nontrivial. Our numerical estimates indicate that using, for example, a proximitized semiconductor nanowire \cite{Mourik2012,Krogstrup2015,Gazibegovic2017,Kang2017,Sestoft2018,Krizek2019} and a periodic arrangement of potential gates is highly unlikely to generate a strong-enough periodic potential. 

\begin{figure}[t]
\begin{center}
\includegraphics[width=.48\textwidth,height = 4.cm]{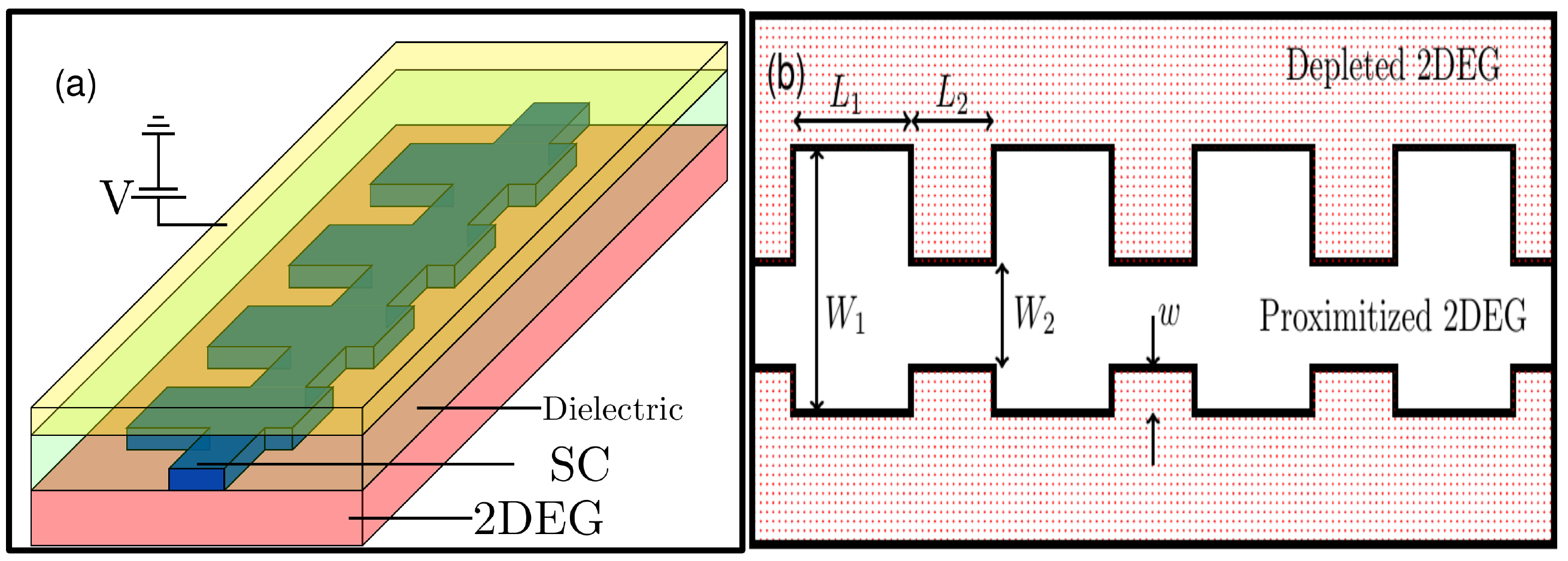}
\end{center}
\vspace{-5mm}
\caption{(a) Schematic representation of the proposed {\em Majorana waveguide} device. A 2DEG hosted by a semiconductor heterostructure  is proximitized by a quasi-1D superconductor with alternating wide and narrow  regions. A top gate depletes the 2DEG outside the region covered by the  superconductor generating a quasi-1D channel with periodically modulated width. (b) Top view of the periodic quasi-1D channel.}
\label{FIG_2_1}
\vspace{-3mm}
\end{figure}

In this section we approach the critical problem of realizing the periodic potential from a different angle: instead of actually applying an {\em external} periodic  potential, we propose the realization of an {\em effective} periodic potential by modulating the width of the device. More specifically, we propose the realization of periodic Majorana devices based on quasi-1D channels realized in 2D semiconductor heterostructures proximity coupled to superconductor strips of periodically modulated width. The feasibility of such a device is supported by recent progress in engineering Al-InAs two-dimensional heterostructures showing experimental signatures consistent with Majorana physics \cite{Shabani2016,Suominen2017,Nichele2017,Fornieri2019,Lee2019,Mayer2019}. 
A schematic representation of the proposed device is shown in Fig. \ref{FIG_2_1}(a). A semiconductor quantum well hosting a two-dimensional electron gas (2DEG) with large spin-orbit coupling and g-factor is proximity-coupled to a conventional superconductor (e.g., Al) grown on top of the semiconductor heterostructure. The structure is capped by a top gate that can deplete the 2DEG  outside the region covered by the superconductor, while the area under the superconductor is screened. As the width of the superconductor is periodically modulated, we obtain a quasi-1D channel with periodic position-dependent width. A top view of the periodic quasi-1D channel is shown in Fig. \ref{FIG_2_1}(b). As shown explicitly below, the periodic modulation of the superconductor width generates a Majorana wire with effective periodic potential  basically equivalent to the 1D model investigated in Sec. \ref{1D}. 

We model the proposed Majorana waveguide by considering a 2DEG with Rashba spin-orbit coupling and position-dependent confining potential described by the  Hamiltonian 
\begin{equation}
\begin{aligned}
    H_{2DEG} =& 
    \left[
    -\frac{\hbar^2}{2m^*}\left(\partial_x^2 + \partial_y^2\right)
    - \mu + V\left(x,y\right)
    \right] \sigma_o \\
    &+ \alpha_y k_x \sigma_y - \alpha_x k_y \sigma_x,
\end{aligned}    
\end{equation}
where $m^*$ is the effective mass, $\mu$ is the chemical potential, $\alpha_x$ and $\alpha_y$ are Rashba coefficients, and $V\left(x,y\right)$ is a  confining potential periodic in the $x$-direction with period $\ell$, $V\left(x+\ell,y\right) = V\left(x,y\right)$. Let us first consider a ``hard-wall'' confining potential, 
\begin{equation}
    V\left(x,y\right) = 
    \begin{cases}
    0, & \text{proximitized region in Fig. \ref{FIG_2_1}(b)}, \\
    \infty, & \text{depleted region in Fig. \ref{FIG_2_1}(b)}.
    \end{cases}
\end{equation}
The width of the proximitized region is $x$-dependent, alternating between $W_1$ and $W_2$, with a bottom offset $w$, as shown in Fig.  \ref{FIG_2_1}(b). First, we study the spectrum of an infinite channel by imposing periodic boundary conditions in the $x$-direction. In the limit $W_1 = W_2$, $w = 0$, and $\alpha_x = 0$, the spectrum is trivially given by the analytic expression
\begin{equation}
\begin{aligned}
    E_{n,p,\tau}\left(k_x\right) =& 
    \frac{\hbar^2}{2m^*} 
    \left( \frac{\pi^2 n^2}{W_1^2} + 
    \left(k_x + \frac{2 \pi p}{\ell}\right)^2\right) \\
    &-\mu + \left(\sigma_z\right)_{\tau,\tau} \alpha_y k_x, 
\end{aligned}    
\end{equation}
where $-\pi / \ell < k_x \leq \pi / \ell$, $n \in \mathbb{Z}^+$, $p \in \mathbb{Z}$, and $\tau = 1,2$. The quantum number $n$ indicates the transverse mode, while $p$ is the zone number, and $\tau$ refers to the $\pm$y spin-$\frac{1}{2}$ eigenstates. Note that the spectrum, which consists of folded and shifted parabolas, is gapless. Allowing $W_1 \neq W_2$ and/or $w \neq 0$ leads to the opening of energy gaps near $k = 0$ and $k = \pm \pi / \ell$. This property can be viewed as a consequence of  plane waves scattering at the interface between the regions with different widths. 
%
\begin{figure}[t]
\begin{center}
\includegraphics[width=.48\textwidth]{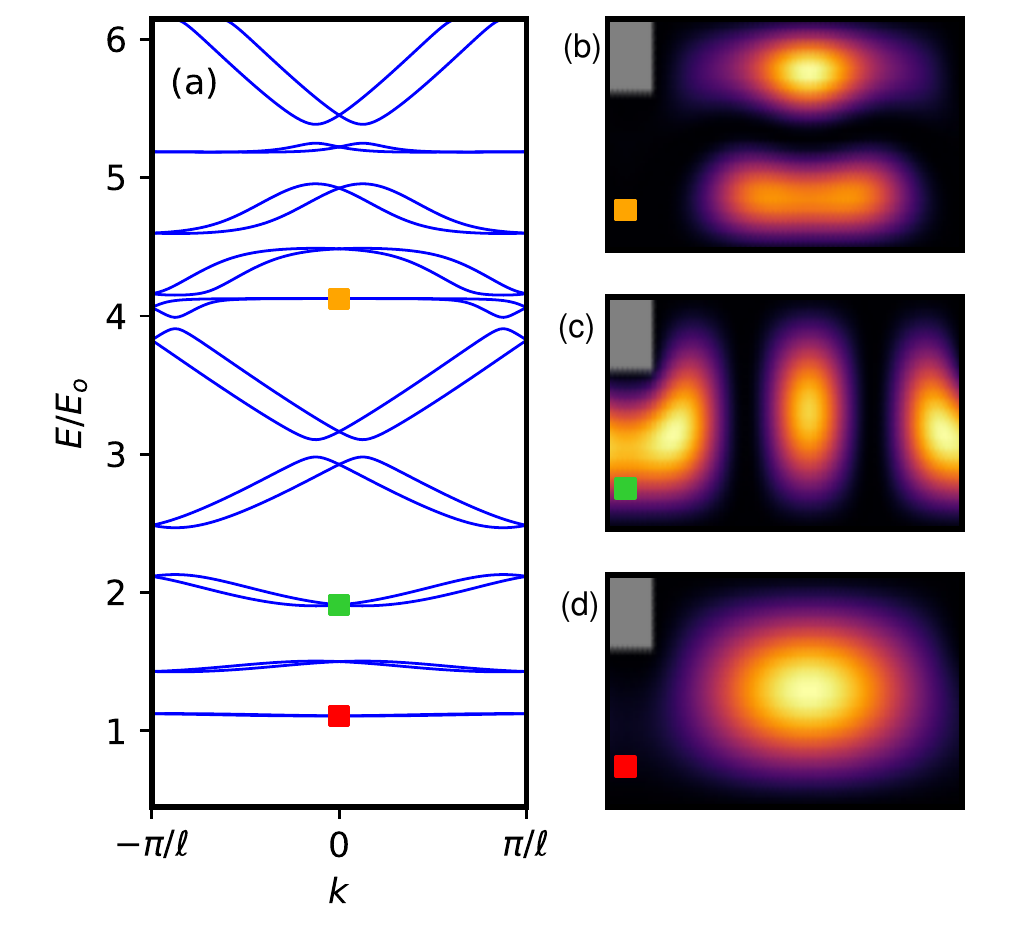}
\end{center}
\vspace{-5mm}
\caption{(a) Spectrum of  a quasi-1D system with ``hard-wall'' confining potential. (b-d)  Modulus square of the $k = 0$ eigenstates marked by colored squares in panel (a). The gray areas outside the narrow regions (of width $W_2$) are inaccessible due to the ``hard-wall'' confinement. The system parameters are: $W_2/W_1 = 0.68$, $w = 0$, $L_1/W_1 = 2.6$, $L_2/W_1 = 0.4$, $m^* \alpha_y W_1 / \hbar^2$ = 0.13, and $\alpha_x =0$, and  the energy unit is  $E_o = \left(\pi^2\hbar^2/2m^*W_1^2\right)$.}
\label{FIG_2_2}
\vspace{-3mm}
\end{figure}
%
The calculated spectrum of a system with $W_1 \neq W_2$ and  $w = 0$ is shown in Fig. \ref{FIG_2_2} (a). Since the eigenenergies scale as $E \propto W_1^{-2}$, as long as the ratios between the spatial variables are held fixed \cite{Weisshaar1991}, it is natural to define the energy unit as $E_o = \left(\pi^2\hbar^2/2m^*W_1^2\right)$, i.e., the confinement energy associated with the first transverse mode in a quasi-1D channel of width $W_1$.
Notice that the first miniband has its bottom just above $E_o$. The (modulus squared of the) wave function corresponding to the lowest energy state [marked by a red square in  Fig. \ref{FIG_2_2} (a)] is shown in  Fig. \ref{FIG_2_2} (b). Notice that the wave function has a single transverse lobe localized within the wide region and does not leak significantly into the narrow region. The minimum energy required to have an oscillatory component within the narrow region is $E = E_o (W_1/W_2)^2 \approx 2.16 E_o$. 
The wavefunction corresponding to the $k = 0$  state of the third minibabd is shown in Fig. \ref{FIG_2_2}(c). The state is still dominated by the first transverse mode, but it has three maxima within the wide region. States characterized by transverse modes with two (or more) maxima ocur above $E \approx 4 E_o$, as expected based on the fact that the confinement energy of the second transverse mode in a uniform channel is $4 E_o$.  An example of such states is show in Fig. \ref{FIG_2_2} (d).
Similarly to the first miniband, this miniband is quite flat, as the second transverse mode decays in the narrow regions for energies below $E \approx 8.65 E_o$. However, this miniband mixes with other minibands (dominated by the first transverse mode) near the zone edge, where it acquires some dispersion. In general, minibands above $4 E_o$ are characterized by strong mixing between different transverse modes. Below this threshold energy, however, the system has a spectrum  similar to that of a purely 1D system in the presence of a periodic potential, which can be seen by comparing the low energy minibands of Fig. \ref{FIG_2_2}(a) and  Fig. \ref{FIG4}(a). Therefore, we expect the results obtain based on the purely 1D model of Sec. \ref{1D} to hold at least for the first few minibands of the quasi-1D structure. 

\begin{figure}[t]
\begin{center}
\includegraphics[width=.48\textwidth]{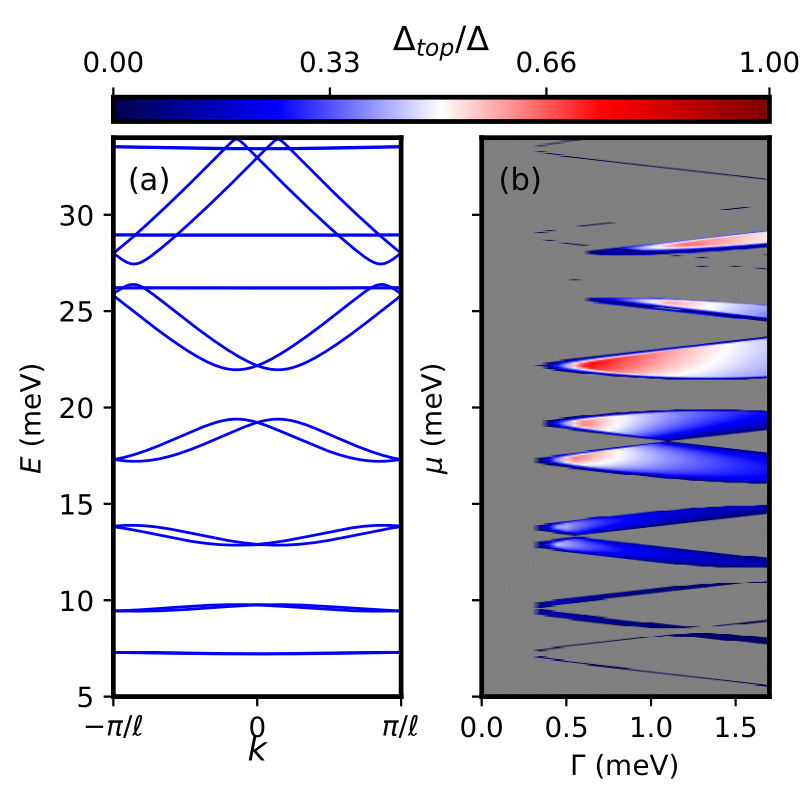}
\end{center}
\vspace{-5mm}
\caption{(a) Spectrum of  a quasi-1D periodic system with mirror-symmetric ``hard-wall'' confining potential. (b) Topological phase diagram of the periodic system as a function of chemical potential, $\mu$, and Zeeman splitting, $\Gamma$. Gray regions are topologically trivial, while colored regions are topologically  non-trivial, with the color scale indicating the size of the topological gap. Note that the phase diagram is qualitatively similar to the phase diagram of the 1D model shown in Fig. \ref{FIG4}(b). The system parameters are: $m^* = 0.023~ m_o$, $\alpha_x = \alpha_y = 100~\text{meV} \cdot \text{\AA}$, $\Delta = 0.3~\text{meV}$, $W_1 = 50~\text{nm}$, $W_2 = 30~\text{nm}$, $L_1 = 130~\text{nm}$, $L_2 = 20~\text{nm}$, and $w = \left(W_1-W_2\right)/2 = 10~\text{nm}$.}
\label{FIG_2_3}
\vspace{-3mm}
\end{figure}

Next, we investigate numerically the correspondence between the properties of the Majorana waveguide and the predictions of the 1D toy model. 
For concreteness, we consider a quasi-1D system with effective parameters consistent with an InAs/Al heterostructure implementation of the proposed device:
 $m^* = 0.023~ m_o$, $\alpha_x = \alpha_y = 100~\text{meV} \cdot \text{\AA}$, $\Delta = 0.3~\text{meV}$, $W_1 = 50~\text{nm}$, $W_2 = 30~\text{nm}$, $L_1 = 130~\text{nm}$, $L_2 = 20~\text{nm}$, and $w = \left(W_1-W_2\right)/2 = 10~\text{nm}$. 
 The corresponding normal spectrum for $\mu = \Gamma= 0$ is shown in Fig. \ref{FIG_2_3}(a). Note the qualitative similarity 
 with the spectrum in Fig. \ref{FIG_2_2}(a). The absence of mixing between the flat and the dispersive high-energy minibands is due to a mirror symmetry  with respect to the $x$-axis generated by the specific choice of geometric parameters. The corresponding topological phase diagram as a function of  Zeeman field, $\Gamma$, and chemical potential, $\mu$, is shown in Fig. \ref{FIG_2_3}(b). The topologically trivial and non-trivial regions correspond to the gray and colored regions, respectively, with the color scale indicating the size of the topological gap. 
The phase diagram displays the main qualitative features found in the context of the 1D model (see Fig. \ref{FIG4}). Again,  each miniband edge supports a (low-field)  topological phase region, with the flat minibands generating a non-trivial phase only inside thin slices of parameter space. Also, the topological gap generally increases as the chemical potential moves into higher energy minibands, due to a larger effective spin-orbit coupling. Note, however, the effect of  higher energy flat minibands (absent in the 1D model), which can create narrow  topological regions  at higher values of the chemical potential.
In practice, these regions are likely irrelevant, due to their small areas (requiring a high degree of fine tuning), and, most importantly, due to their small topological gaps. 

\begin{figure}[t]
\begin{center}
\includegraphics[width=.48\textwidth]{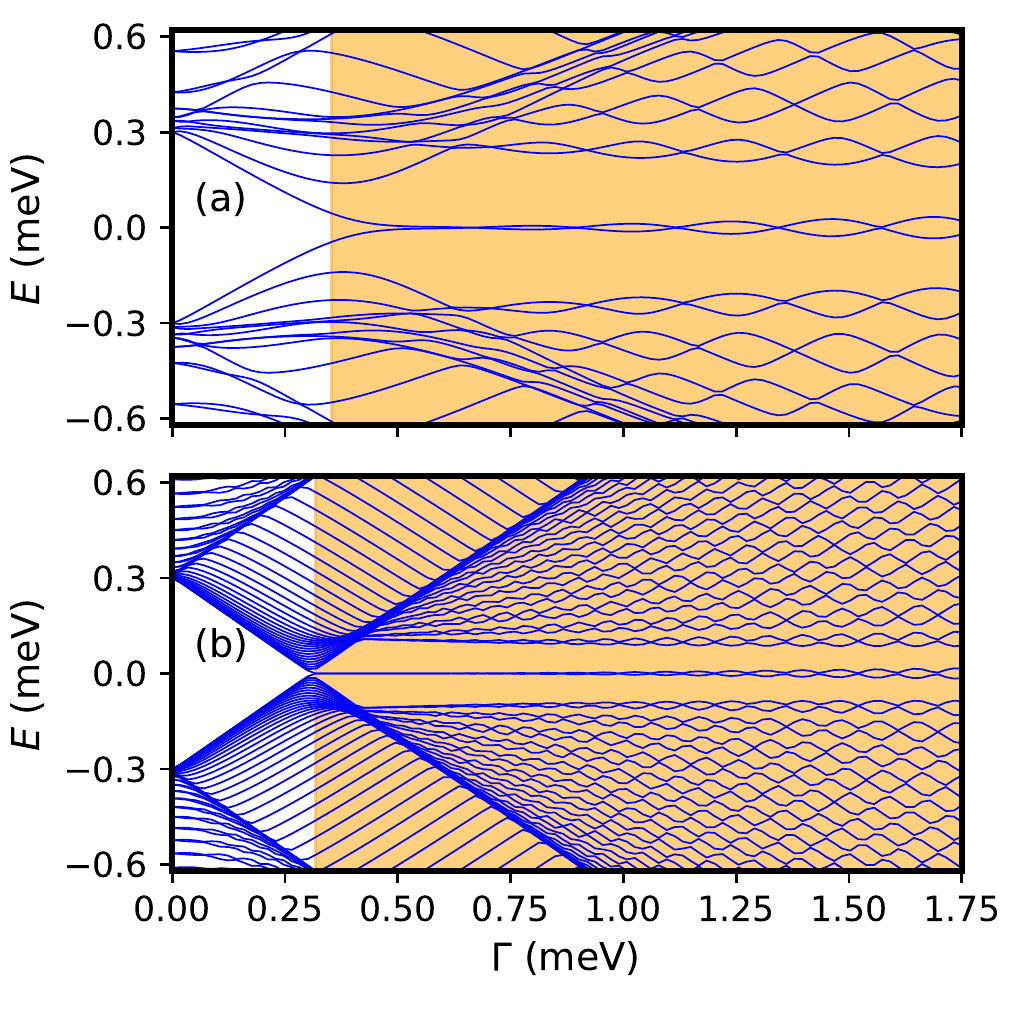}
\end{center}
\vspace{-5mm}
\caption{(a) Low energy spectrum of a finite  quasi-1D periodic system as a function of the Zeeman field $\Gamma$. The length of the system is $L = 3.75~\mu\text{m}$ and the chemical potential is $\mu = 22.2~\text{meV}$. All other parameters are the same as in  Fig.\ref{FIG_2_3}.  (b) Spectrum of a uniform system of thickness $W_1 = 50~\text{nm}$ (same as the wide region of the periodic system) and chemical potential corresponding to the bottom of the first transverse mode. All other parameters are the same as in (a).}
\label{FIG_2_4}
\vspace{-3mm}
\end{figure}

Consider now a finite segment of length  $L = 3.75~\mu\text{m}$ of the  quasi-1D system discused above. The dependence of the low-energy spectrum on the applied Zeeman field for  $\mu = 22.2~\text{meV}$ is shown in Fig. \ref{FIG_2_4}(a).  For comparison,  the spectrum of a uniform wire with constant thickness $W_1 = 50~\text{nm}$ and the chemical potential tuned to the bottom of the first transverse mode (all other parameters being the same) is shown  in Fig. \ref{FIG_2_4}(b). As expected based on our 1D analysis, the periodic system with the chemical potential within a higher energy miniband supports a larger topological gap and has a lower density of states, compared to the uniform system. Also note that  the Majorana mode in Fig. \ref{FIG_2_4}(a) is characterized by larger energy splitting oscillations as compared to its counterpart in the uniform system [see Fig. \ref{FIG_2_4}(b)], indicating  a larger MBS localization length, in agreement with the results of Sec. \ref{1D}.

\begin{figure}[t]
\begin{center}
\includegraphics[width=.48\textwidth]{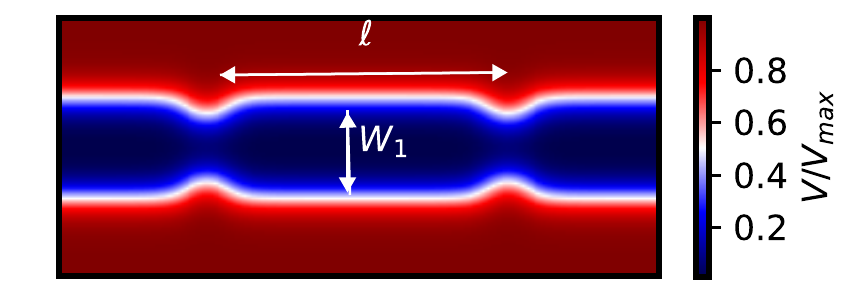}
\end{center}
\vspace{-5mm}
\caption{Confining potential landscape for a smooth potential defined by Eq. (\ref{sPot}) with parameters: $W_1 = 50~\text{nm}$, $W_2 = 30~\text{nm}$, $L_1 = 130~\text{nm}$, $L_2 = 20~\text{nm}$, and $\chi = 10~\text{nm}$. }
\label{FIG_2_5}
\vspace{-3mm}
\end{figure}

While the ``hard-wall'' potential is a convenient approximation for initially exploring the physics of Majorana waveguides, in real systems there is a finite length scale, $\chi$, over which the confining potential varies from its value  inside the covered region to its value inside the depleted region. 
 In fact, it is important that this ``soft-confinement'' length scale be large-enough, as this enables the tuning of the chemical potential by the top gate.
 Quantitatively, the  length scale $\chi$, which describes the efficiency of the screening by the superconductor of the potential created by the top gate, depends on the details of the heterostructure and can be estimated by solving a numerically challenging Schrodinger-Poisson problem \cite{Woods2018,Antipov2018,Mikkelsen2018}.  Here, we do not address this problem, but instead consider $\chi$ as a phenomenological parameter and investigate the following key question:  how are the topological phase diagram and the properties of the Majorana bound states affected by a finite (rather than zero) screening length? To address this question, we model the ``smooth'' confining potential as 
\begin{equation}
    V\left(x,y\right) = V_{max}
    \left[
    g(x) V_1(y) + \left(1-g(x)\right) V_2(y)
    \right], \label{sPot}
\end{equation}
where
\begin{align}
    g(x) =& 
    H(x,0,\chi) - H(x,L_1,\chi), \\
    V_1(y) =& 
    H(-y,0,\chi) + H(y,W_1,\chi), \\
    V_2(y) =& 
    H(-y,w,\chi) + H(y,w+W_2,\chi),
\end{align}
with $H$ defined in Eq. (\ref{Heavi}). An example of a smooth confining potential landscape is shown in Fig. \ref{FIG_2_5}. Note that the finite screening length $\chi$ rounds the corners of the wide and narrow regions, resulting in a smooth periodic quasi-1D channel. 

\begin{figure}[t]
\begin{center}
\includegraphics[width=.48\textwidth]{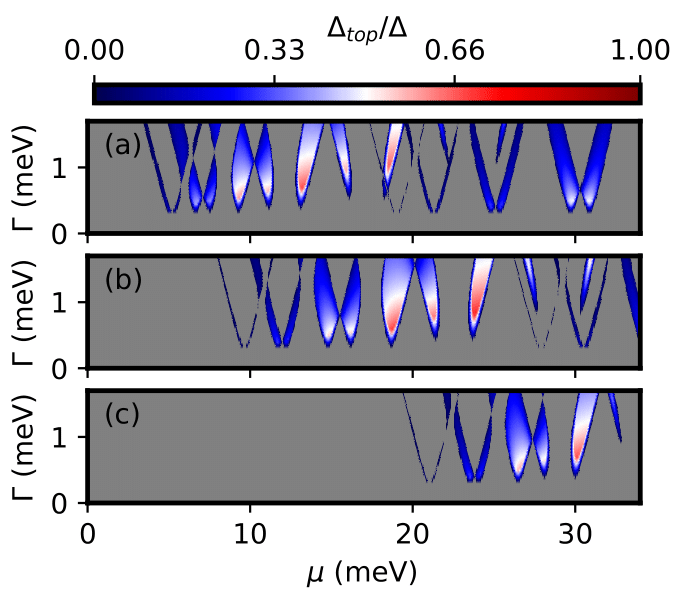}
\end{center}
\vspace{-5mm}
\caption{Topological phase diagram for a system with soft-confinement. The confining potential is given by  Eq. (\ref{sPot}), with  $V_{max} = 75~\text{meV}$ and different screening lengths: (a) $\chi = 0$, (b) $\chi =10~\text{nm}$, and  (c) $\chi =20~\text{nm}$.
Other system parameters are the same as in Fig. \ref{FIG_2_4}. Increasing the screening length $\chi$ shifts phase diagram toward larger values of $\mu$, but does not change the main features.}
\label{FIG_2_6}
\vspace{-3mm}
\end{figure}

The dependence of the phase diagram on the screening length $\chi$ is illustrated in Fig. \ref{FIG_2_6}. The system is characterized by a soft confinement potential given by Eq. (\ref{sPot}) with  $V_{max} = 75~\text{meV}$ and different values of the screening length. All other parameters are  the same parameters as in Fig. \ref{FIG_2_4}. In the top panel, i.e.,  Fig. \ref{FIG_2_6}(a), we have $\chi = 0$, which means a sharp transition from the covered region to the depleted region, similar to the hard-wall scenario, except for a finite value of the potential in the depleted region.
The resulting  phase diagram is practically identical to the hard-wall phase diagram in Fig. \ref{FIG_2_4}(b), except for an overall shift  to smaller values of the chemical potential due to a (small) leakage of the wave functions into the depleted region. As the screening length $\chi$ increases, the topological regions move to higher values of $\mu$ as the proximitized (i.e., covered) channel acquires a non-zero potential.
Importantly, the phase diagram remains largely unaffected by the finite $\chi$, even for values of the screening length comparable to the width of the channel [e.g., in Fig. \ref{FIG_2_6}(c) we have $\chi = 0.67W_2 = 0.4W_1$]. 
Of course,  the exact shape and location of the phase boundaries change slightly with $\chi$, but the overall topological area, the lowest values of the critical Zeeman field , and the typical size of the topological gap are practically insensitive to changes of the screening length. This insensitivity, combined with the overall shift of the phase diagram, demonstrates the possibility of tuning the chemical potential using the top gate (i.e., changing $V_{max}$) without altering the phase boundaries. For the parameters used in these calculations, a variation of the chemical potential on the order of $5~$meV (i.e. $10\Delta$) practically guarantees access to a miniband edge, which corresponds to a low-field topological phase. In addition, the overall energy scale can be further controlled through the channel geometry, in particular the length scales $L_1$, $L_2$, $W_1$, and $W_2$ that determine the confining potential landscape. Having demonstrated the basic equivalence between the low-energy physics of the Majorana waveguide and the effective 1D model of Sec. \ref{1D}, we conclude that periodic  quasi-1D channels engineered using patterned 2D semiconductor-superconductor structures represent an extremely versatile platform that can provide significant advantages for realizing robust Majorana zero modes.  

\section{Summary and conclusions}

In this work we investigate the emergence of topological superconductivity and Majorana zero modes in periodic structures, focusing on (i) determining the impact of the periodic potential on the robustness of the Majorana modes and (ii) identifying practical solutions for engineering a strong-enough periodic potential. First, we consider a simplified one-dimensional model of the periodic structure and derive analytical expressions for  the  effective parameters that characterize the periodic potential-induced minibands. We find that higher energy minibands are characterized by large values of the effective spin-orbit coupling and low effective masses. In turn, this leads to enhanced values of the topological gap and low densities of bulk states, creating ideal conditions for the realization of robust topolgical superconductivity. Using numerical simulations, we show explicitly that the presence of a periodic potential reduces the low-field parameter space consistent with topologically trivial low-energy states that mimic the local phenomenology of Majorana zero modes (i.e., ps-ABSs or quasi-Majoranas). Most importantly, the periodic potential enhances the robustness of the Majorana modes against disorder. To demonstrate this point, we introduce an edge-to-edge correlation function that could prove useful in determining the expected output of non-local (three terminal) conductance measurements on Majorana devices. The price for the larger topological gap and the increased robustness against disorder is an enhanced Majorana  localization length characterizing the periodic structures. While this implies longer wires for realizing topologically protected Majorana zero modes,  in the near term this property (combined with the enhanced robustness against disorder and the reduced susceptibility of generating trivial low-energy states) can prove helpful in demonstrating hybridization-induced Majorana oscillations, a key feature that, so far, has eluded experimental observation, casting doubts regarding the real nature of the observed zero-bias conductance peaks. In addition, the enhancement of the  Majorana  localization length is rather moderate (a factor of two, or less), provided the periodic potential is strong-enough, a key condition for actualizing the potential benefits of periodic structures.


To address the critical problem of engineering strong-enough periodic potentials, we propose the  \textit{Majorana waveguide} device, a quasi-1D channel with periodically modulated width hosted by a semiconductor heterostructure and proximity-coupled to a lithographycally defined superconductor.
We show that the periodic modulation of the channel width can generate an effective periodic potential with the required specifications. Moreover, 
in the regime characterized  by a chemical chemical potential smaller than the confinement energy of the second transverse mode, the phase diagram of the Majorana waveguide has all the qualitative features predicted by the 1D effective theory (compare, for example,  Figs. \ref{FIG4} and \ref{FIG_2_4}). 
Finally, we show that the finite length scale associated with the screening by the superconductor has a weak effect on the phase boundaries, basically resulting in an overall shift of the chemical potential. This property suggests the possibility of efficient control of the chemical potential by a top gate.  In addition, the effective system parameters can be engineered by controlling the channel geometry, in particular the length scales $L_1$, $L_2$, $W_1$, and $W_2$ that determine the confining potential landscape. These elements, which support the feasibility of conditions necessary for actualizing the potential benefits of periodic structures identified by our analysis of the effective 1D model, suggest that modulated quasi-1D channels realized in patterned 2D semiconductor-superconductor structures provide a promising platform for realizing robust Majorana zero modes.

Future theoretical efforts that could assist the fabrication and measurement of this type of devices should build on detailed Schr\"{o}dinger-Poisson simulations of realistic structures. Obtaining quantitative estimates of the screening length will enable the optimization of the channel geometry and the  identification of the realistic range for the top gate lever arm. In addition, based on the calculated screening length, one can estimate the strength of the effective disorder generated by patterning  imperfections, which addresses the important engineering problem regarding how precise these patterns need to be.  Finally, these efforts must be supplemented by calculations of transport properties, including quantitative predictions of  edge-to-edge correlations 
expected in a multi-terminal measurement.

\appendix
\section{Localization of Majorana bound states} \label{LOCAL}

As we did in the main text, we consider the 1D Rashba nanowire in an applied magnetic field and with s-wave superconductivity, expect we include no periodic potential.
The Hamiltonian is given by 
\begin{equation}
H = \left(\frac{\hbar^2k^2}{2m^*} - \mu + \alpha k \sigma_y + \Gamma \sigma_z \right)\tau_z  - \Delta \sigma_y \tau_y, \label{Ham1}
\end{equation}
where $\mu$ is the chemical potential, $\Gamma$ is the Zeeman energy, $\Delta$ is the induced superconductivity parameter, and $\sigma_i$ and $\tau_i$ are the Pauli matrices acting in spin and particle-hole space, respectively. This Hamiltonian is known to undergo a topological phase transition at a Zeeman field given by 
\begin{equation}
    \Gamma_c = \sqrt{\Delta^2 + \mu^2},
\end{equation}
with Majorana bound states emerging at the edges of the wire once this critical magnetic field is reached. While the Majorana states will have zero energy for a semi-infinite system, the overlap between the edge Majorana causes oscillations about zero energy as $\Gamma$ is changed. This overlap can be quantified if we know the length scale on which the Majorana bound states are localized at the edge of the wire. Since the Majorana bound state's energy lies inside the bulk energy bands, the states are necessarily composed of evanescent waves that have a complex wave number. The imaginary component of the complex wavenumber causes decay of the edge mode as it enters into the bulk of the wire. We can study the length scale of this decay by studying the complex band structure of Eq. (\ref{Ham1}), where we let $k$ become a complex number.

To begin we find the eigenenergies of Eq. (\ref{Ham1}), which are found to be
\begin{equation}
\begin{aligned}
    E^2(k)& = \left(\frac{\hbar^2k^2}{2m^*} - \mu\right)^2 + \Gamma^2 + \Delta^2 + \alpha^2 k^2 \\
    &\pm 2\sqrt{ \left(\frac{\hbar^2k^2}{2m^*} - \mu\right)^2  (\Gamma^2 + \alpha^2 k^2) + \Gamma^2 \Delta^2}. \label{Eigs}
\end{aligned}    
\end{equation}
Note that for any eigen energy $E$, there as also exists and eigen energy $(-E)$ due to the particle-hole symmetry of the Eq. (\ref{Ham1}). Since Majorana bound states are (nearly) zero energy modes, we desire to find $k$ satifying Eq. (\ref{Eigs}) for $E = 0$. In principle, this can be done exactly since finding $E=0$ solutions to Eq. (\ref{Eigs}) involves solving a quartic equation for the variable $k^2$. The quartic equation solution is too complicated to be of practical use, however, so we instead approach the problem using asymptotic methods \cite{Bender1999}. Note that in the limit of $\Delta = 0$, the Hamiltonian (\ref{Ham1}) separates into particle and hole components. The spectrum of the isolated particle sector is simply
\begin{equation}
    E(k) = \frac{\hbar^2k^2}{2m^*} - \mu \pm \sqrt{\Gamma^2 + \alpha^2 k^2},
\end{equation}
and we can easily find the $E = 0$ solutions, which are 
\begin{equation}
\begin{aligned}
    k_o^2 &= \frac{2 m^*}{\hbar^2}\left(\mu + 2 E_{SO}\right) \\
    &\pm \frac{2 m^*}{\hbar^2}\sqrt{\Gamma^2 + 4\mu E_{SO} + 4E_{SO}^2}, \label{ko}
\end{aligned}    
\end{equation}
where $E_{SO} = m^* \alpha^2 / \left(2 \hbar^2\right)$
is the spin-orbit energy. Note that $k_o$ represents an asymptotic approximation for $k$ in Eq. (\ref{Eigs}) as $\Delta \rightarrow 0$. To find the leading order correction, we let $k^2 = k_o^2 + \left(2 m^*/\hbar^2\right) \Delta p^2$ and substitute this expression into Eq. (\ref{Eigs}). Here $p^2$ is a dimensionless parameter that we wish to find. A question arises as to what sign to take in Eq. (\ref{ko}). To answer this question, notice that for large enough $\Gamma$, we obtain a set of purely real and a set of purely imaginary values for $k_o$. The purely real and imaginary sets of eigen values correspond to the low and high energy spin split bands, respectively, where the high energy spin band has imaginary wavenumbers because no propagating bulk states exist at $E = 0$ for this band. In the presence of $\Delta \neq 0$, the real eigenvalues are rotated slightly into the complex plane, while the imaginary eigenvalues remain completely imaginary. The localization length of the Majorana modes will be determined by the eigenvalues with the smallest imaginary component. Therefore, the eigenvalues stemming from the real eigenvalues of Eq. (\ref{ko}) are the most important, so we take the ($+$) sign. 
\begin{figure}[t]
\begin{center}
\includegraphics[width=0.48\textwidth]{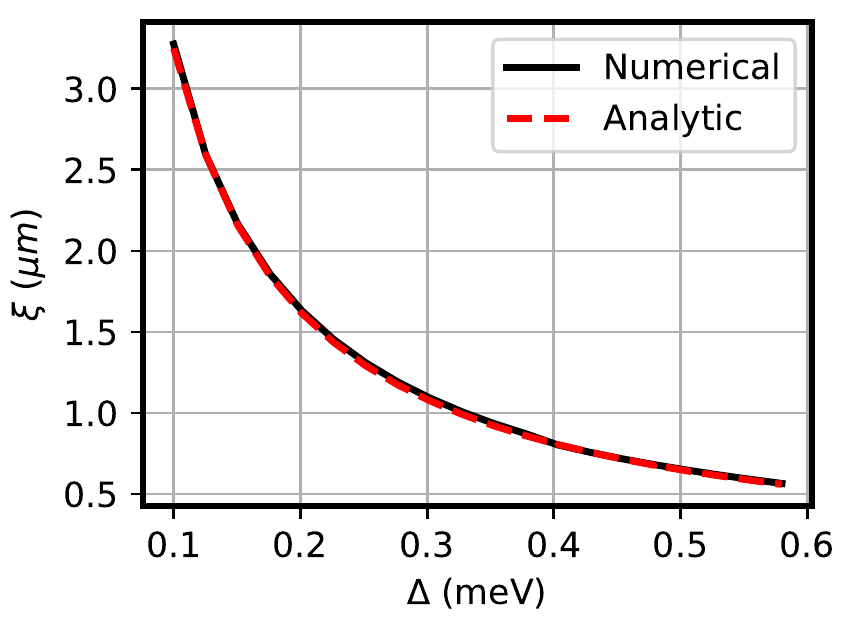}
\end{center}
\vspace{-5mm}
\caption{Localization length, $\xi$, vs superconducting gap, $\Delta$, for a long nanowire, $\xi \ll L$, with parameters $m^* = 0.026 m_o$, $\mu = 1$ meV, $\alpha = 50~\text{meV}\cdot~\text{nm}$, and $\Gamma = 3$ meV. Numerical localization length is extracted by fitting the Majorana wavefunction to the exponential envelope, $\psi \propto$ exp$\left(-x/\xi\right)$.}
\label{FIG_A1}
\vspace{-3mm}
\end{figure}
Upon substituion, we find the asymptotic relation,
\begin{equation}
p^2 \sim \pm i \sqrt{ \frac{4 E_{SO} k_o^2}{\Gamma^2 + 4 \mu E_{SO} + 4 E_{SO}^2}},
\end{equation}
as $\Delta \rightarrow 0$. Our total wave number is then given by
\begin{align}
k &= \pm \kappa \pm i q,\label{wavenumber} \\ 
\kappa &\sim k_o \\
q &\sim \sqrt{ \left(\frac{2m^*}{\hbar^2}\right) \frac{E_{SO}}{\Gamma^2 + 4\mu E_{SO} + 4 E_{SO}^2}} \Delta,
\end{align}
where the two $\left(\pm\right)$ in Eq. (\ref{wavenumber}) are independent. The localization length is then just the reciprocal of the imaginary part of the wave number, $\xi = q^{-1}$. We find \begin{equation}
\xi \sim \ell_{SO} \left( \frac{\Gamma}{\Delta} \right)\sqrt{1 + \frac{4\mu E_{SO} }{\Gamma^2} + \frac{4E_{SO}^2}{\Gamma^2}},    \label {Analy}
\end{equation}
where $\ell_{SO} = \hbar^2/\left(m^* \alpha\right)$ is the spin-orbit length. This shows the expected behavior of increased delocalization for large $\Gamma$ and small $\alpha$.

Interestingly, in the limit of $E_{SO} \gg \Gamma$, we actually find increasing localization length for increasing $\alpha$. While this regime is difficult to achieve for the conventional proximitized Rashba nanowire setups, this situation is possible in the higher-energy minibands within a periodic potential setup. A comparison between the numerical and derived analytical localization length of Eq. (\ref{Analy}) is shown in Fig. \ref{FIG_A1} as a function of $\Delta$, showing excellent agreement. We note that starting from the assumption of a small spin-orbit coefficient $\alpha$ in Eq. (\ref{Eigs}), we find a similar and consistent asymptotic expression for the localization length, given by 
\begin{equation}
    \xi \sim \ell_{SO} \frac{\sqrt{\Gamma^2 - \Delta^2}}{\Delta},
\end{equation}
as $\alpha \rightarrow 0$ in the case of $\mu = 0$.




%

\end{document}